\documentclass[preprint]{aastex}

\shorttitle{Djorgovski et al.}
\shortauthors{Collapsed Cores in Globular Clusters}

\begin{document}


\title{Infrared Colors and Variability\\ of Evolved Stars\\ from 
COBE DIRBE Data}


\author{Beverly J. Smith}
\affil{Department of Physics and Astronomy, East Tennessee State University,
    Box 70652, Johnson City, TN  37614}
\email{smithbj@etsu.edu}



\begin{abstract}

For a complete 12 $\mu$m
flux-limited sample of 207
IRAS sources (F$_{12}$ $\ge$ 150 Jy, $|$b$|$ $\ge$
5$^{\circ}$), the majority of which are AGB stars
($\sim$87$\%$), 
we have extracted 
light curves in 
seven infrared bands between
1.25 $\mu$m $-$ 60 $\mu$m using
the 
database of the 
Diffuse Infrared Background Experiment
(DIRBE) instrument on the Cosmic Background Explorer (COBE) satellite.
Using previous infrared surveys, we filtered these light curves
to remove datapoints affected by nearby companions, and obtained time-averaged
flux densities and infrared colors, as well as estimates of their variability
at each wavelength.
In the time-averaged DIRBE color-color plots,
we find clear segregation
of 
semi-regulars, Mira variables,
carbon stars, OH/IR
stars, and 
red giants without circumstellar dust
(i.e., V $-$ $[$12$]$ $<$ 5) and with
little or no visual
variation ($\Delta$V $<$ 0.1 magnitudes).
The DIRBE 
1.25 $-$ 25 $\mu$m 
colors become progressively
redder and the variability in the DIRBE database
increases along the oxygen-rich sequence 
non-dusty slightly varying red giants 
$-$$>$
SRb/Lb $-$$>$ SRa $-$$>$ Mira $-$$>$ OH/IR 
and the carbon-rich SRb/Lb $-$$>$ Mira sequence.
This supports previous assertions that these are evolutionary
sequences involving the continued production and ejection of dust.
The carbon stars
are redder than their oxygen-rich counterparts for the same
variability type, except in the F$_{12}$/F$_{25}$ ratio, where they
are bluer.
We find significantly larger infrared variations for Mira
stars, carbon stars, and OH/IR stars than for 
the red giants without circumstellar dust, with 
semi-regulars and irregulars in between, 
as expected based on their optical variations.
Of the 28 sources in the sample not previous
noted to be variable, 18 are clearly variable in the DIRBE data,
with amplitudes of variation of $\sim$0.9 magnitudes at 4.9 $\mu$m
and $\sim$0.6 magnitudes at 12 $\mu$m, consistent with them being
very dusty Mira-like variables.
We also present individual DIRBE light curves of a few selected
stars.
The DIRBE light curves of 
the semi-regular
variable L$_2$ Pup are particularly remarkable.  The maxima at 1.25, 2.2, and 3.5 $\mu$m occur
10 $-$ 20 days before those at 4.9 and 12 $\mu$m, and, at 4.9 and 12 $\mu$m,
another maximum is seen between the two near-infrared maxima.

\end{abstract}


\keywords{stars: AGB and post-AGB, stars: variable: Miras }

\section{INTRODUCTION AND MOTIVATION}

\subsection{Asymptotic Giant Branch Stars and Infrared Emission}

Asymptotic giant branch (AGB) stars 
are in the last
stage of stellar evolution before becoming planetary
nebulae, and are generally surrounded by circumstellar dust shells.   
AGB stars typically have high rates of mass loss, 
between 10$^{-8}$ $-$ 10$^{-4}$ M$_{\sun}$ yr$^{-1}$ \citep{km85},
sometimes losing more than half their mass while on the AGB.
These stars
return large quantities of carbon-, oxygen-, and nitrogen-enriched
gas to the interstellar medium, contributing significantly
to the chemical evolution of the Galaxy \citep{s94}.
AGB stars include classical
Mira variables (visual amplitudes $\ge$ 2.5; periods $\ge$ 100 days),
semiregular SRa stars (amplitudes $\le$ 2.5; periods 35 $-$ 1200 days),
semiregular SRb stars (amplitudes $\le$ 2.5; with poorly defined periods),
and
irregular variables (type Lb), as well as
dust-enshrouded
carbon stars and OH/IR stars.
AGB stars can be divided into two main groups depending upon
their chemistry, oxygen-rich and carbon-rich.
Stars begin their lives on the AGB as oxygen-rich and some, but
not all, later become carbon-rich because of carbon dredge-up
from their core \citep{ir83,
ck90}.

To better understand the mass loss process
in AGB stars and the evolutionary relationship
between the different types of AGB stars, mid- and far-infrared studies 
are vital.  At wavelengths longer than $\sim$3~$\mu$m, 
emission from dust in the circumstellar envelopes of AGB stars
becomes important \citep{wn69}.
Infrared observations
have proven useful in constraining models of the 
circumstellar
envelopes of these stars, as well as estimating mass loss rates
(e.g., \citet{j86}, \citet{b87}; \citet{apv93}).
Such studies 
have provided important clues to the dust properties, temperatures,
and distributions in these envelopes.

A common method of 
distinguishing between AGB stars and other sources
is segregation in 
IRAS 
color-color plots 
(e.g., \citet{t87},
\citet{vdvh88},
\citet{ie00}).
When IRAS data are combined with shorter wavelength
infrared data, 
this segregation is improved
and 
it is possible to separate the different types of AGB stars.
For example,
\citet{ik00}
were able to separate SRb/Lb stars from 
Miras, and  
oxygen-rich AGB stars from carbon-rich AGB stars,
by their location in the 25 $\mu$m$-$12 $\mu$m vs. 12~$\mu$m-K
color-color plot.
Adding 3.5 $\mu$m data improves the segregation: the K $-$ L vs.
[12] $-$ [25] and K $-$ L vs. L $-$ [12]
color-color diagrams are particularly
useful in segregating different types of objects \citep{e87,
g93}.
Such infrared color-color plots have provided clues to 
possible
evolutionary connections between the different types of AGB stars
(e.g., 
\citet{kh92},
\citet{vdvh88}, \citet{vdv89},
\citet{wj88},
\citet{ck90},
\citet{j90}, \citet{mik01}).

Many of the early studies of the circumstellar shells
of AGB stars
were based on time-averaged
mid- and far-infrared data
(for example, \citet{j86}, \citet{apv93}).
A large fraction of AGB stars, however, are 
variable at visible and near-infrared wavelengths,
and may well be variable at longer wavelengths as well, causing
some uncertainty in these results.
Measurements of variability at longer infrared wavelengths provide much
stronger constraints of these models.
For example, a number of studies have shown that, for 
Mira variables,
the amplitude of variation tends to decrease with increasing
wavelength, from the optical to the mid-infrared \citep{lw71,
h74, lb92, lb93, lss96, smith02}.
Comparisons of mid-infrared light curves with those
in the
visible and near-infrared provide information
about how the circumstellar shell changes in response to
changes in the star.  
Interesting phenomena such as wavelength-dependent phase
lags, where the light curve maximum occurs at different times
at different wavelengths, and secondary maxima in the light
curves, can provide important additional constraints on models of 
AGB stars and their
circumstellar dust shells.

\subsection{The COBE DIRBE Mission and Infrared Light Curves of Variable Stars}

Most 
mid-infrared light curves obtained to date are relatively
poorly-sampled, making measurements of amplitudes,
phase lags, and secondary maxima difficult.
To address these issues, in \citet{smith02},
we recently explored the use
of 
archival 
data from the 
Diffuse Infrared Background Experiment (DIRBE) \citep{h98}
on the Cosmic
Background Explorer (COBE) satellite 
\citep{b92}
for extracting infrared light curves of Mira variable stars.
DIRBE operated for 10 months in 1989-1990, providing full-sky
coverage at 10 infrared wavelengths
(1.25, 2.2, 3.5, 4.9, 12, 25, 60, 100, 140, and 240 $\mu$m).
Although DIRBE was designed to 
study the cosmic infrared background, its data are also useful for
studying point sources, in spite of its relatively 
poor spatial resolution (0.7$^{\circ}$).
To date, however,
stellar fluxes from DIRBE have been little
utilized, except for our Mira study and for calibration verification
during the DIRBE mission \citep{bm97, 
c98}.  

DIRBE 
provided excellent temporal
coverage, allowing careful investigation of possible 
infrared variability
of stars.
In the course of a week, a typical point on the sky was observed 10-15 times
by DIRBE,
and over the course of the cryogenic mission, it was
observed about 200 times \citep{h98}.
The DIRBE coverage varied with position on the sky. Near the ecliptic
poles, sources were generally observed approximately twice a day for the entire mission,
giving a total of 400 $-$ 1000 observations over the 10 month mission.
Near the ecliptic plane, sources were typically observed
roughly twice a day for 2 months, then
were inaccessible for 4 months, before coming back into view.
For comparison, the Infrared Astronomical
Satellite (IRAS) typically provided only two or three independent
flux measurements of a star at 12 $\mu$m 
\citep{ls92},
while the infrared
Midcourse Space Experiment (MSX) mission \citep{p99}
made up to six observations
over a four month period \citep{e99}.

In our earlier study \citep{smith02},
we started with the samples of \citet{sp98}
and \citet{slp98}.
These samples were
obtained by cross-referencing the General Catalogue
of Variable Stars \citep{GCVS} with the \citet{PSC}
and selecting sources
with IRAS Low Resolution Spectroscopy available.
We then selected a subset of 38 stars
from this sample, with an IRAS 12 $\mu$m flux
density cutoff of 235 Jy and classification as Mira stars.
This study proved that high quality 1.25 $\mu$m $-$ 25 $\mu$m
light curves of infrared-bright stars are obtainable
from the DIRBE dataset. 
In general, the highest
S/N light curves are at 3.5 $\mu$m and 4.9 $\mu$m, with decreasing
S/N at longer wavelengths.
This study confirmed and extended previous results on decreasing amplitudes
of variation
with increasing wavelengths.
In addition, 
we 
found evidence for phase lags of 
the near-infrared maximum relative
to that in the visual,
and offsets in the times of the mid-infrared maxima relative
to those in the near-infrared.  The mid-infrared
maxima occurred $\sim$0.05 phase before those in the near-infrared, but
$\sim$0.05 phase after those in the optical.  
In addition, in three stars,
clear secondary maxima in the rising portions
of their light curves are seen
simultaneously
in the optical, near-infrared, and mid-infrared, 
supporting the hypothesis that they are due to shocks rather than
newly formed dust layers.

In our earlier study, we specifically targeted AGB stars
that were previously known to be Mira variables.  
This means that we neglected non-Mira AGB stars (semi-regular variables
and irregular variables), Miras
not previously classified as Miras, and
very dust-enshrouded 
AGB stars without bright optical counterparts.
In our current study, we have extended this work to these other types of
AGB stars and have increased the sample size to 207.
In this paper, we discuss the time-averaged DIRBE 
colors of the different types of AGB stars, as well as statistical
results on 
infrared variability.  

\section{The Sample and Some Statistics }

For this new study,
instead of selecting our stars from the \citet{sp98}
and \citet{slp98} samples, which were specifically
chosen to include only known variables, we have
started with a more general infrared-selected sample of stars.
For DIRBE variability studies, the best wavelength to use
would be 4.9 $\mu$m or 3.5 $\mu$m, since these give the highest
S/N DIRBE light curves for AGB stars \citep{smith02}.
Unfortunately, however,
no complete all-sky survey has been made at these wavelengths
so far.
We therefore use the IRAS 12 $\mu$m band to select our sources,
selecting all IRAS sources with F$_{12}$ $\ge$ 150 Jy
and $|$b$|$ $\ge$ 5$^{\circ}$.
This is the average
5$\sigma$ 12 $\mu$m noise level in the light curves
obtained from the DIRBE Calibrated Individual Observations (CIO) database
\citep{smith02}.

The Galactic latitude limit was chosen to minimize confusion because of
the large
DIRBE beam (0.7$^{\circ}$).
At the DIRBE 12 $\mu$m 5$\sigma$ limit of 150 Jy,
the IRAS source counts show that
there is on average 0.065 sources per square degree at 
$|$b$|$ = 5$^{\circ}$.
This corresponds to 28 DIRBE beams per source and a confusion noise
of 28 Jy, giving a S/N of 5 at 150 Jy.
Thus at higher flux limits and higher Galactic latitudes, confusion
is expected to be less important.

Our full sample contains 207 sources.
All but six of the sources in the \citet{smith02}
study are included
in the new sample; the remaining six did not meet the 
Galactic latitude requirement.
Our sample is listed in 
Table 1\footnote{Table 1 is only available 
in
electronic form, 
at the Astronomical Journal
web site http://www.journals.uchicago.edu/AJ/ or at
http://www.etsu.edu/physics/bsmith/dirbe/},
along with information
about possible optical counterparts to the IRAS sources from
the SIMBAD database \citep{w96} and \citet{kvb97},
including
optical spectral type and the IRAS spectral classification.
We also identify stars known to be OH/IR stars 
with 1612 MHz OH satellite line emission
(from the compilation of \citet{c01}),
as well as objects identified as HII regions, planetary nebulae, and post-AGB
stars in SIMBAD.
Of the 207 sources in our sample, 
29 are in the \citet{c01}
compilation of
OH/IR sources.

In Table 1, we also include information about the variability
type, when available.
At least 179 of the sources in the sample (86$\%$) are known 
or suspected from previous
studies to be variable.  Cross-correlating with 
the GCVS \citep{GCVS},
we find 158 matches.  An additional 20 sources not
in the GCVS are included in the New Catalogue of Suspected Variable
Stars \citep{NSVS}
or the NSV Supplement \citep{kd98}.
Some statistics on the variability types of the sample sources
are given in Table 2, where the variability type is compared to 
optical type.
Of the previously-known variable sources in the sample,
82 are classified as Miras, 7 are 
semi-regular type SRa, 39 are SRb stars, 6 are SRc stars, three are unspecified
semi-regulars, 15 are type Lb, two are Lc stars,
one is an unspecified type L star,
one is a symbiotic
Z And-type+semi-regular binary pair, one is a possible type R variable (binary
with strong reflection of light from hot star
on companion), one is
an eclipsing binary, and one is a suspected
irregular eruptive variables of type IN (Orion variable;
likely a pre-main sequence object).
The rest are unclassified variables.

All 207 sources in the sample are
in the \citet{kvb97} IRAS Low Resolution Spectra (LRS) catalog
sample.  The IRAS spectral types given in Table 1 are from \citet{kvb97},
who
classified the IRAS 
spectra according
to the scheme of \citet{vc89}.
Some statistics on these IRAS spectral types are summarized in Tables 3 and 4, where
they are compared with the optical spectral types and the variability types.
The IRAS types indicate that at least 171 (82$\%$) of our
207
sample sources are likely dusty evolved
stars (LRS types A, C, E, or F).  This includes 46 (22$\%$ of the sample)
carbon-rich stars (type C, with strong 11.2 $\mu$m SiC emission),
106 (51$\%$) oxygen-rich stars with 9.7 $\mu$m silicate emission
(type E), one oxygen-rich star 
with the silicate feature in absorption (type A), and 18 (9$\%$) type F stars
(featureless IRAS spectra that are flatter than expected from
a stellar photosphere, indicating small amounts of dust).
In addition, 27 sources (13$\%$) have class S LRS spectra, meaning a Rayleigh-Jeans
photospheric spectrum with little to no dust.
Finally, four sources are type H (generally planetary nebulae, reflection
nebulae, or HII regions), one object is type I (noisy or incomplete spectrum),
two are type P (red 13 $-$ 23 $\mu$m spectra, and a sharp rise at
the blue end of the LRS range),
while two are type U (unusual spectra that do not fit the other classes).

As noted by \citet{kvb97} and as
seen in Table 3, the optical spectral type (carbon vs. oxygen-rich star) is strongly
correlated with the dust spectral type, as seen in the IRAS
mid-infrared 
Low Resolution Spectra (LRS) database, in that carbon
stars tend to show 11 $\mu$m SiC emission, while
M-type AGB stars tend to have the 9.7 $\mu$m silicate feature.
The LRS type A star is an M star, as are
all of our LRS type E stars, with 
the exception of two optical type S stars
(C/O $\sim$ 1)
and 11 stars with no optical types.
Of the M stars with LRS type E, two 
are luminosity class II stars and three are SRc variables, thus 
are likely massive stars rather than AGB stars.
The LRS type C stars either have no optical spectral type or are
carbon stars, with the exception of one 
M star.
The LRS type F stars are all class M stars (including one M supergiant),
except for one K giant.
Of the LRS type S stars, 5 are supergiants, one is a carbon star, 
and 21 are 
giants.
Three of the four LRS class H sources are HII regions or are associated
with star formation
regions, and the other is a planetary nebula.
The LRS type I source is a semi-regular M giant, while one of the 
LRS class U stars is a carbon Mira and the other is a K supergiant.
One of the LRS class P sources is a post-AGB star while the other is associated
with a star formation region.

In Table 1, we also include the V magnitude range, if available from
one of the three variable star catalogues cited above.  Otherwise, if possible, we include
the V magnitude given in SIMBAD.  A total of 155 of our 207 sources 
have V magnitudes available.
In Figure 1, a histogram of V $-$ $[$12$]$ for these 155 stars is given.
The shaded region marks the 27 stars with IRAS spectral types of S (dominated
by photospheric emission).  As expected, these are the bluest stars 
in the sample in V $-$ $[$12$]$.
For comparison, note that a cool M star with a blackbody spectrum
is expected to have V $-$ $[$12$]$ $\le$ 7 \citep{c87}.
Note, however, that almost all of the LRS type S stars
in our sample are known from previous studies to
be variable (see Table 4), although none have the large amplitudes that characterize
Miras.
A star at our 12 $\mu$m flux limit with V = 10 has a V~$-$~$[$12$]$ = 11.8;
thus this plot is likely incomplete at V $-$ $[$12$]$ higher than this.

To search for trends in the infrared colors and variability of 
the different types of AGB stars,
we have divided our sample stars into 26 groups (see Table 6), 
separating according to the 
different variability types (Miras,
SRa, SRb, SRc, Lb, Lc, sources noted only as `variable' in
the GCVS or NSV, and sources not previously classified
as variable).  We distinguished 
between
carbon stars, oxygen-rich stars, and S stars, and 
separated out sources known to be
OH/IR stars.
We also separated out known non-carbon luminosity class I and II stars, since
these are likely to be massive stars, and therefore not on the AGB.
Luminosity classes and/or variability types are not available for all of the stars
in the current study, however, since supergiants are rare
and main sequence stars have low luminosities, we assume
that if this information is not available the star is probably
a giant star.

The non-supergiant oxygen-rich sources
with unspecified variability are an inhomogeneous
group, containing both 
optically-faint stars without many optical or near-infrared measurements,
and
well-studied 
optically-bright stars 
with small variations.
To separate these two categories of stars, we further sub-divided this group
by V $-$ $[$12$]$.   There are six 
optically-bright (V $<$ 3.5)
very blue 
(V $-$ $[$12$]$ $<$ 5)
K and M giants
in this group
which are slightly variable at optical wavelengths ($\sim$0.1 magnitudes in V).
In contrast, the other five unspecified non-supergiant variables are very red, with 
V $-$ $[$12$]$ $>$ 13.5, or have no V magnitudes available (i.e., 
V $-$ $[$12$]$ likely $>$ 12).
These two groups are separated in Table 6, since they are likely in different evolutionary
stages.
In addition, there is one optically-bright (V = 2.23) very blue star 
(V $-$ $[$12$]$ $\sim$ 4.1) non-variable star, $\gamma$ Dra,
in the oxygen-rich stars
without variability information group.  This star is in marked contrast to the rest of the oxygen-rich
stars with no variability information, which have no V magnitudes available.  
Thus $\gamma$ Dra is separated
out into a different group.
Also, the single known eclipsing binary in the sample, 
the optically-bright (V = 1.9) star $\epsilon$ Car, 
which contains a K3~III star, is
very blue 
(V $-$ $[$12$]$ $\sim$ 4.25) with
an LRS spectral type of S, and shows no signs of variability in the DIRBE data.  Thus
this is grouped with $\gamma$ Dra.
These eight optically-bright non-variables and slightly variable stars are 
likely not AGB stars, but rather red giant branch or red clump stars.

Of the 207 sources in our full sample, 180
(87$\%$)
are likely AGB stars, 
based on their optical and IRAS spectral types.
Of the remainder, four are associated with 
star formation regions, 13 are non-carbon supergiants, bright giants, and/or SRc variables 
(and
therefore likely massive stars), one is a post-AGB star, one is a planetary
nebula, and 8 are likely giant branch or red clump stars.

\section{The COBE Data}

In Table 5,
the time-averaged DIRBE flux densities in the six
shortest wavelength DIRBE bands are given\footnote{Table 5 is only available
in electronic form, at the Astronomical
Journal web site at http://www.journals.uchicago.edu/AJ or at
http://www.etsu.edu/physics/bsmith/dirbe/},
These values were extracted from the DIRBE CIO database using
software written by Nils Odegard.  The CIO database contains the
calibrated individual 1/8th second samples taken in science-survey
mode during each day of the DIRBE cryogenic mission.
For all scans that pass within 0.3$^{\circ}$ of the target position,
a linear baseline is fit to the sections 
$\pm$1.35$^{\circ}$$-$2.25$^{\circ}$
from the point
of closest approach.  The point source photometry is obtained
by subtraction of this baseline and correcting for DIRBE beam response.
The uncertainties in the point source photometry are calculated as
the quadrature sum of the rms noise of the baseline,
an error due to positional uncertainties
of 1$'$ in both the in-scan and cross-scan directions,
an error due to short-term detector gain variations, and
signal-dependent detector noise.
The average noise levels per individual flux measurements
are 25, 20, 20, 10, 30, 55, 320, 765, 4800, and 2750 Jy
for the DIRBE 1.25 $-$ 240 $\mu$m filters, respectively.

To test for possible contributions to the 
DIRBE 12, 25, and 60 $\mu$m flux densities from infrared-bright stars near
the targeted sources, we have searched the
IRAS Point Source Catalog
for additional infrared-bright sources 
within 0.5$^{\circ}$ of the target source.
At 12, 25, and 60 $\mu$m, 4, 2, and 2 sources in our list,
respectively, had nearby sources within 0.5$^{\circ}$ above the COBE
1$\sigma$
noise levels for individual flux measurements given above.
These objects are flagged in Table 5. 
The two sources flagged at 25 and 60 $\mu$m are both star formation
regions, one of which is also flagged at 12 $\mu$m.

As an additional test for confusion, 
for our sample sources
we compared the time-averaged 
DIRBE
12, 25, and 60 $\mu$m flux densities 
with the
IRAS Point Source 
Catalog flux densities.
Greater than 3$\sigma$ 
disagreements were found for 
5, 24, and 13 sources at 12, 25, and 60 $\mu$m, respectively.
These sources are also flagged in Table 5.
These flagged objects include all four
of the H~II regions in the sample as well as the planetary
nebula.  
A few evolved stars are also flagged.
A few of these flagged evolved stars
have been found to be extended
at 60 and 100 $\mu$m in the IRAS data
\citep{hawkins89, hawkins90a, hawkins90b,
ypk93a, ypk93b}, however,
in all but the case of $\alpha$ Ori the DIRBE flux densities are
significantly
larger than the total IRAS flux densities, suggesting additional sources
of confusion.

At 1.25 and 2.2 $\mu$m, 
we use the all-sky
Two Micron All Sky Survey 
(2MASS)\footnote{http://www.ipac.caltech.edu/2mass/} 
\citep{c03} to check for additional
bright sources in the DIRBE beam.
At 1.25 and 2.2 $\mu$m, respectively, 6 and 58 sources were
flagged as having a companion 
within 0.5$^{\circ}$
above the DIRBE noise level.
At 3.5 and 4.9 $\mu$m,
no all-sky surveys are available.  At these wavelengths,
to check for possible additional sources in the DIRBE beam,
we used the Catalog of Infrared Observations
\citep{gps99}\footnote{Available from the
VizieR service at
http://vizier.u-strasbg.fr/viz-bin/VizieR},
a compilation of all published infrared observations
available in 1997.
At 4.9 $\mu$m, 
we also used 
the synthetic all-sky 4.2 $\mu$m catalog of \citet{ep96},
which was created by extrapolation from the IRAS catalogues,
other infrared catalogues,
and optical measurements.
A total of 3 sources were flagged as having companions above the 
DIRBE noise levels at 3.5 $\mu$m in the DIRBE beam, while 24 sources
were flagged at 4.9 $\mu$m.
We note that these catalogs 
may be incomplete, even at these very
bright limits, thus these
are lower limits to the numbers of confused sources at these wavelengths.

Another issue
is the possibility of more distant companions
affecting the `sky' fluxes used in the CIO photometry.
Careful inspection of the DIRBE data showed that, 
if a second infrared-bright star is
between about 
0.5$^{\circ}$ $-$ 2.5$^{\circ}$ of the target star, then
for scans that passed through the nearby star,
flux from the companion star
sometimes contributed significantly to the `sky flux'
used in the DIRBE point source photometry routine,
causing erroneous photometry with large errorbars for the targeted star. 
Fortunately, however, scans in other directions were not affected by the 
second star.
To correct for this problem,
we filtered our data to remove affected scans.
At 12, 25, and 60 $\mu$m, we searched the IRAS Point Source
Catalog for objects within 3.2$^{\circ}$ of each targeted source.
At each wavelength, for each DIRBE scan for each targeted source, 
we scaled the IRAS flux densities
of the nearby stars by a Gaussian with
FWHM 0.7$^{\circ}$, weighted by the minimum distance between the
scan and the companion star.
If the weighted IRAS flux density in the respective
band was greater than the DIRBE noise limit, then the scan was removed
from consideration.
This technique dramatically improved the DIRBE light curves of
some stars, removing discrepant data points and those with large errorbars.
At 12, 25, and 60 $\mu$m, respectively, 
71, 24, and 7 of our sources had some scans affected by
a bright companion. 
We note that some uncertainty still remains,
since this process does not take into account the possibility
that the infrared brightnesses
of some of the nearby stars may change with time.

In a few cases, the nearby sources are extended, and, although
the source clearly affects some of the DIRBE scans, the
Point Source Catalogue fluxes are below our cut-off.
To test for these cases, we also searched in the IRAS 
Small Scale
Structure Catalog \citep{hw85}
for companions above our 
flux cutoff, and filtered the DIRBE data accordingly.
At 12, 25, and 60 $\mu$m, only 1, 10, and 11 
sources respectively were affected by such companions.

At 1.25 and 2.2 $\mu$m,
we used the 2MASS database to filter the data, while at 3.5 $\mu$m,
we used 
the Catalog of Infrared Observations, and at
4.9 $\mu$m, we used both the Catalog of Infrared Observations
and the \citet{ep96} catalog.
At 1.25 $\mu$m, 2.2 $\mu$m, 3.5 $\mu$m,
and 4.9 $\mu$m, 82, 205, 56, and 193 out of our 207 sources had at least some scans
potentially affected by a nearby companion.

In some cases, this filtering dramatically improved the light curves
(see Figure 2).
However,
because of incompleteness in the comparison
catalogs and the fact that some of the nearby stars may 
themselves be variable,
this filtering is sometimes not perfect.  
Also, 
at our brightness levels, the 2MASS observations
were saturated, so the photometry is relatively inaccurate 
($\sim$0.2 $-$ 0.3 magnitudes; \citet{c03}), which
introduces some additional uncertainty into
the filtering process.
We therefore
inspected each final filtered light curve by eye, 
looking for obvious evidence for companions
(as in Figure 2).  These light curves were also flagged in Table 5.

These filtering routines cannot correct for cosmic ray
hits, which can also cause anomalous points in the DIRBE
light curves.  The remaining datapoints with large
error bars ($\ge$3 times the average uncertainty)
were likely affected by a cosmic ray hit in the `sky' portion
of the light curve, and so were removed from the light curves by
an additional filtering process.
Generally, only a few DIRBE datapoints per light curve
were removed by this additional
filtering processing.  

We note that a cosmic ray hit near the star itself may not cause
a large error bar, but instead may simply manifest itself as a very
discrepant flux measurement with a small
error bar.  The typical time resolution of the DIRBE data
is $\sim$9 hours, thus a true rapid variation of
the star would show the same signature.
Large amplitude visual variations (0.5 $-$ 1 magnitude) with timescales
of hours or days have been claimed on the basis of Hipparcos
data \citep{dl98} (but see \citet{kll01}).
To look for such events in the DIRBE data, we compared the flux in a single
measurement with the mean of the closest six measurements in time.
Points that differed by greater than 5$\sigma$ were then 
inspected individually.
In none of the unconfused sources did such a large deviation occur at 
three or more 
DIRBE wavelength simultaneously, strongly indicating that these
points are caused by cosmic rays rather than real variations.
We therefore removed these points from the light curves
for the following variability analysis.

In addition to the time-averaged infrared flux densities, 
Table 5 also
includes both the standard deviation of the individual flux values in the 
light curve of the object (after filtering),
along with the mean uncertainty of the individual
data points in the light curve.  The comparison of these two values
provides an estimate of the likelihood
of variability of the object (See Section 5.2).  The
standard error of the mean is also provided, along with
the number of data points (sightings) for the objects after the final
filtering.
In addition, the observed amplitude of variation at each wavelength
(see Section 4.2)
is also included in Table 5.

No color corrections have been applied to the data in Table 5.
To investigate the uncertainties associated with color corrections,
we created lookup tables of expected observed flux density ratios for adjacent DIRBE
bands for a large range of blackbody temperatures, using
$\lambda$$^0$, $\lambda$$^{-1}$, and $\lambda$$^{-2}$ emissivity laws
and integrating over the DIRBE responsivities.
Comparison with the observed DIRBE flux ratios for the sample stars
yielded equivalent temperatures.  For each DIRBE band except 1.25 $\mu$m
and 25 $\mu$m, the average of the
two temperatures obtained from comparison with the two adjacent bands 
were 
used with
the DIRBE color correction tables \citep{h98} 
to estimate color corrections.   
At 1.25 $\mu$m and 25 $\mu$m, only the temperatures implied by the 1.25 $\mu$m/2.2 $\mu$m 
and 12 $\mu$m/25 $\mu$m ratios, respectively, were used.
At 3.5, 4.9, 12, and 25 $\mu$m, the implied corrections were quite small,
typically
$\le$5$\%$, 
differing by a few percent depending upon the assumed emissivity
law. 
At 
1.25 $\mu$m and 2.2 $\mu$m the implied color corrections
were somewhat larger, up to
$\sim$10$\%$, differing by up to $\sim$5$\%$ with the different emissivity laws.  
Since the color corrections are small,
and the uncertainty associated with the emissivity
law is on the same order as the color corrections themselves, we elected to
not include color corrections in the following analysis.

In Figures 3 and 4, we show some representative final filtered
light curves of different types
of stars.  These stars are discussed below.  

\section{RESULTS}

\subsection{Time-Averaged Infrared Color-Color Plots}

In Figures 5 $-$ 8, we present the 
time-averaged DIRBE color-color diagrams for the stars in our sample, using
the six shortest wavelength DIRBE bands.
In these plots, we have
excluded
sources flagged in Table 5 at at least one of the relevant wavelengths.
We also excluded
the star formation regions, planetary nebulae, post-AGB
sources, and the symbiotic binary.
For clarity,
we have also excluded sources likely to be massive stars
(luminosity classes I and II, variability types SRc and Lc).

In Table 6, we provide the mean colors and rms 
spread for our 26 groups 
of stars.
These results are summarized in Figures 9 and 10, where we plot
the average colors for the different types of AGB stars, along
with their dispersions.
For comparison with these plots and tables, note that 
J $-$ K = $-$2.5 log (F$_{1.25}$/F$_{2.2}$) + 0.98 
and K $-$ L = $-$2.5 log(F$_{2.2}$/F$_{3.5}$) + 0.89 \citep{bb88},
while $[$12$]$ $-$ $[$25$]$ = $-$2.5 log (F$_{12}$/F$_{25}$) + 1.56
\citep{b88}, though filter differences and color corrections 
may cause slight differences.

These 
color-color plots, along with Table 6, show a clear segregation 
of different types of
objects.  The semi-regulars are bluer than Miras on average, 
with Miras known to be OH/IR stars being redder than
oxygen-rich Miras not classified as OH/IR stars.  
Carbon-rich
Miras are redder than oxygen-rich Miras in general, and
carbon-rich stars not previously classified as Miras are also very red.
Among the semi-regulars, the SRa stars are redder than the SRb stars.
Among the known OH/IR stars, there is a trend, in that the known
semi-regulars are bluest, objects without a specified variability type
are reddest, and Miras are in between.
The bluest colors belong to 
the optically-bright red giants with little or no
visual variations (for example, $\beta$ And, 
$\alpha$ Hya, $\alpha$ Boo, and $\beta$ UMi), all of which
are IRAS LRS type S (dominated by photospheric emission in the IRAS spectra).
These stars likely 
are not yet AGB stars.

For the 155 stars with V measurements, in Figures 11 and 12, we compare 
V $-$ $[$12$]$ 
with the four shortest wavelength DIRBE colors.
Note that these plots are likely incomplete above 
V $-$ $[$12$]$ $\sim$ 11.8. 
As expected, in most cases, the DIRBE colors get redder
as 
V~$-$~$[$12$]$ gets redder.  However, note that for
the oxygen-rich stars F$_{3.5}$/F$_{4.9}$ gets bluer
as 
V~$-$~$[$12$]$ 
reddens to 
V~$-$~$[$12$]$ $\sim$ 7, then the trend reverses.
Note also that, for a given
V $-$ $[$12$]$,
the carbon stars are redder in
F$_{1.25}$/F$_{2.2}$ and F$_{2.2}$/F$_{3.5}$ than the oxygen-rich stars.  
In the V $-$ $[$12$]$ vs.
F$_{3.5}$/F$_{4.9}$ and F$_{12}$/F$_{25}$ plots,
the carbon and oxygen-rich stars are indistinguishable.

\subsection{DIRBE Variability}

For each light curve with minimum flux density greater than
five times the typical DIRBE noise level for that wavelength,
we have calculated the 
total
{\it observed} change in brightness during
the DIRBE observations, after averaging over
one week time intervals.
We note that these observed changes in brightness do not necessarily
represent
the full range of variation for these stars,
because many of the light curves are not complete and may not cover
a full pulsation period.  In these cases, the observed variations
are a lower limit to the true amplitudes of variation.
Of the unflagged sources, excluding sources with less than
5$\sigma$ DIRBE flux densities at minimum light, 
at 1.25, 2.2, 3.5, 4.9, 12, and 25 $\mu$m, 
respectively,
81, 107, 103, 138, 96, and 47 stars have $>$3$\sigma$ detections of variability
(i.e., $\Delta$(mag)/$\sigma$$_{\Delta(mag)}$ $>$ 3),
and 
57, 89, 79, 106, 66, and 28
stars have $\ge$5$\sigma$
detections of variability.
One star, the carbon star CW Leo (IRC~+10216), has greater than 3$\sigma$
variation at 60 $\mu$m and 100 $\mu$m.  It varies by 
1.63 $\pm$ 0.05,
1.47 $\pm$ 0.04, 
1.17 $\pm$ 0.01, 0.82 $\pm$ 0.02, 0.67 $\pm$ 0.02, 0.49 $\pm$ 0.05, and 0.43 $\pm$ 0.07 magnitudes
at 2.2, 3.5, 4.9, 12, 25, 60, and 100 $\mu$m, respectively (see Figures 3A $-$ E).

In Figures 13 $-$ 18, we present histograms of these observed variations,
in magnitudes, for the 26 groups of stars in our sample.
The average observed amplitude of variation at each wavelength, 
$<$$\Delta$(mag)$>$,
in magnitudes, for these 26 groups,
is given in Table 7, along with the rms spread.
To demonstrate the significance of 
$<$$\Delta$(mag)$>$, in Table 7 we also include the average
uncertainty in $\Delta$(mag), $<$$\sigma$($\Delta$(mag))$>$.
In contructing Table 7 and Figures 13 $-$ 18, we only included stars with $>$5$\sigma$ DIRBE
flux densities at minimum light.

Miras show significantly larger variations
in the DIRBE data
than the semi-regulars, and the SRa's have larger variations
than the SRb's.
The Miras have average observed
amplitudes of $\sim$1 magnitude at 1.25 $\mu$m and $\sim$0.6
magnitudes at 4.9 $\mu$m, while the non-massive SRa's and SRb's
have average observed amplitudes of 
0.41 and 0.17 magnitudes at 1.25 $\mu$m and 0.15 and 0.09 
magnitudes at 4.9 $\mu$m, respectively.

In Figures 3F $-$ P and 4A $-$ B, we show some example light curves for a few Miras and
semi-regular stars.
Note the inflection point in the rising portion of the
4.9 $\mu$m light curves of the Mira variables
T Dra and R Vol (Figures 3F and 3H).
Such inflection points have been noted before in the mid-infrared
light curves of Miras \citep{smith02}.
Note also the striking difference between the 1.25 $\mu$m and 4.9 $\mu$m
light curves
for the oxygen-rich SRb star L$_2$ Pup.
The maxima at 1.25 $\mu$m precede those at 4.9 $\mu$m by 10 $-$ 20 days.
Also, at 4.9 $\mu$m, a secondary peak is seen between the two 1.25 $\mu$m
maxima.  
The 2.2 $\mu$m and 3.5 $\mu$m light curves (not shown) resemble the
1.25 $\mu$m curve, while the 12 $\mu$m curve (not shown) is similar to that
at 4.9 $\mu$m.

The least variable stars in the sample
are the 
optically-bright giants
without circumstellar dust
(V $-$ $[$12$]$ $<$ 5) 
(including both the stars known to vary slightly
in the optical and the stars with no known variability).
These stars have essentially no observed variations in the DIRBE data 
(3$\sigma$ upper limits $<$ 0.03 $-$ 0.1 magnitudes; Table 7; see for example Figure 4C and D).
In contrast, the optically-faint stars 
not previously classified as variable
are quite variable in the DIRBE data, varying by $\sim$0.9 magnitudes at
4.9 $\mu$m, even more than the classified Miras
(no information is available at 1.25 $\mu$m, since
these stars are too faint at 1.25 $\mu$m to
measure variations).
Inspection of these light curves by eye confirms that, of the 28 stars
in our sample not previously known to be variable, 18 are clearly
variable in the DIRBE data (see, for example, Figure 4E $-$ H).
This boosts the number of known variable stars in
our complete 12 $\mu$m flux-limited sample from 179 to 197, or 95$\%$
of the sample
(note, however, that some of the catalogued
variables are not observably variable in the DIRBE database).
This difference in the observed DIRBE amplitudes of variation of 
the optically-bright stars (V $-$ $[$12$]$ $<$ 5) and the 
optically-faint stars (V $-$ $[$12$]$ $>$ 12) in the combined
`variable'/`not known to be variable' class confirms that these
are two distinct populations of stars.

Among the variability types, there is a slight tendency for
the carbon-rich stars to be more variable than the oxygen-rich
stars (Table 7), however, these differences are small and may
not be statistically significant. 

To investigate how the amplitudes of variation vary
with wavelength, in Figure 19a we plot the {\it observed} 
amplitude at 1.25 $\mu$m against that at 2.2 $\mu$m,
including only stars with 
5$\sigma$ fluxes at minimum brightness at both wavelengths.
In Figures 19b, 19c, 19d, and 19e,
we plot
$\Delta$mag(2.2 $\mu$m) vs. 
$\Delta$mag(3.5 $\mu$m),
$\Delta$mag(3.5 $\mu$m) vs. 
$\Delta$mag(4.9 $\mu$m),
$\Delta$mag(4.9 $\mu$m) vs. 
$\Delta$mag(12 $\mu$m),
and
$\Delta$mag(12 $\mu$m) vs. 
$\Delta$mag(25 $\mu$m), respectively.
In all five plots, correlations are seen.
Inspection of these plots shows that
the average amplitude of variation at the longer
wavelength in each panel is less than at the shorter wavelength.
This is quantified in Table 8, where the mean ratios of the amplitudes
at adjacent wavelengths are provided for the different classes of
object in the sample.
This tabulation excludes stars with flux densities 
less than
5$\sigma$ at minimum and with less than a 5$\sigma$
detection of variation at both wavelengths.
Although there is significant scatter in Figure 19,
in most cases, the average amplitude ratios (Table 8) are greater than one,
indicating
that the amplitude of variation generally decreases with increasing wavelength. 
There is a slight tendency for carbon-rich stars to have a larger
4.9 $\mu$m to 12 $\mu$m ratio than oxygen-rich stars; otherwise, there
is little significant difference between the different classes of stars.

Since the DIRBE cryogenic period was relatively short compared
to a typical AGB star pulsation period, for most stars it
was not possible to determine a period from the DIRBE data.
However, for 103 of the likely AGB stars in our sample, periods
are available from the GCVS or the NSV.
For these stars, in Figure 20 we have plotted the period against the 5 shortest wavelength DIRBE colors.
These plots show that, except for F$_{12}$/F$_{25}$, 
as the period increases, the stars get redder.

\section{DISCUSSION}

\subsection{Oxygen-Rich Stars}

Some clear trends are apparent in the time-averaged DIRBE colors 
and the infrared variability of
the AGB stars in this sample (Tables 6 and 7, and Figures 5 $-$ 18).
For oxygen-rich stars on the AGB, the infrared colors become
redder and the observed amplitudes of variation
increase along the sequence 
visually-bright slightly-varying non-dusty giants 
$-$$>$
SRb $-$$>$ SRa
$-$$>$ Mira $-$$>$ Mira OH/IR stars
$-$$>$ OH/IR stars known to be variable but without a variability classification.
At the four shortest wavelengths, the SRb stars are bluer than the SRa's;
from 4.9 $\mu$m to 25 $\mu$m the colors are indistinguishable.
As noted by \citet{kh92}, the V amplitude criteria
for Mira selection ($\Delta$(mag) $>$ 2.5) is not strictly followed
in the GCVS, in that some nearby well-studied stars that exceed this
limit are classed as SRa or SRb rather than Miras because of irregularities
in their light curves.  The intermediate infrared colors of the SRa's
between Miras and SRb's, along with their amplitudes and periods,
lead \citet{kh92, kh94}
to conclude that oxygen-rich SRa stars
are a mixture of Miras and SRb sources, rather than a unique
type of stars.  
In our sample, both of the carbon SRa's and two of the three oxygen-rich
SRa's have V ranges in the GCVS greater than 2.5 magnitudes, suggesting
that they may be misclassified Miras.   We note, however, that their
infrared colors are bluer than the majority of Miras.

The infrared colors of the Lb stars
are similar to those of the SRb stars, as noted
previously by \citet{klh96}.
The DIRBE variabilities of these two classes (Table 7 and Figures 13 $-$ 18)
are also similar.
Based on a Galactic scale height analysis, \citet{jk92}
concluded that irregulars and semi-regulars belong to the same population.
A detailed study of long term light curves of Lb and SRb
stars by \citet{kll01}
showed that they are essentially the
same class of objects, consistent with our results.

The infrared colors of the optically-faint unspecified oxygen-rich variables
in our sample 
(other than the unspecified OH/IR stars and the supergiants/bright giants) are 
between those of the Miras and the semi-regulars, suggesting that they are
a mixture of those two populations.
This conclusion is supported by the distribution of observed amplitudes
in these classes (Figures 13 $-$ 18 and Table 7).

The ten optically-faint oxygen-rich stars without
previous 
variability information 
are very red, similar to the OH/IR Miras and redder than the non-OH/IR Miras.
Visual
inspection of the DIRBE light curves for these sources suggests
that at least some of them are previously unidentified Miras.
Six are clearly strongly variable in the mid-infrared,
with average amplitudes of variation of approximately 0.75 magnitudes
at 4.9 $\mu$m, 0.6 magnitudes at 12 $\mu$m, and 0.4 magnitudes
at 25 $\mu$m (for example, see Figures 4E and F).  These are consistent with the amplitudes
found for the known Miras in the sample.
The fact that these very red stars with colors similar to
those of OH/IR stars are not known 1612 MHz OH/IR masers suggests
that the OH/IR phase may be intermittent (i.e., \citet{l02}) or that not
all very evolved oxygen-rich stars become observable 1612 MHz sources.

In general, for the stars with V magnitudes available, 
as 
V $-$ $[$12$]$ gets redder, the DIRBE colors also get redder.
For the bluest stars (V $-$ $[$12$]$ $<$ 7), 
which have little circumstellar dust, 
the increase in V $-$ $[$12$]$ is due to increasing titanium oxide
absorption in the visible with decreasing photospheric temperature.
For stars with larger V $-$ $[$12$]$, dust absorption also 
depresses the visible light.
Interestingly, for the oxygen-rich
stars, 
for V $-$ $[$12$]$ 
$\le$ 7,
as 
V $-$ $[$12$]$ gets redder, F$_{3.5}$/F$_{4.9}$ gets bluer (Figure 11c).  
This trend
reverses at redder
V $-$ $[$12$]$, where both colors get redder.
This depression in the 4.9 $\mu$m flux density for very cold 
stars without circumstellar dust is probably due to an increase in 
the 4.3 $-$ 4.7 $\mu$m CO
absorption in these stars.
Spectroscopic observations from the Infrared Space Observatory (ISO)
shows that the equivalent width of this absorption band increases strongly
with decreasing temperature for normal stars without circumstellar shells 
\citep{h02}.

Since the DIRBE cryogenic period is shorter than a pulsation period,
we are not able to use the DIRBE light curves to accurately determine
pulsation periods for these stars.  However, by inspection
of the DIRBE light curves by eye, for 3 of these 10 stars, we are able
to estimate a rough lower limit to the pulsation period for these stars.
These lower limits range from 340 to 370 days, showing, as expected,
that these very red stars have long pulsation periods.

These trends in infrared colors and variations
support a scenario of stellar evolution 
in which, as stars evolve up the AGB, their pulsation amplitudes,
mass loss rates, and opacities increase,
and their infrared colors become progressively redder.  
In this picture, they first evolve from semi-regulars to Miras \citep{kh92}
and then, depending upon their chemistry, to either extreme carbon
stars or OH/IR stars
\citep{vdvh88, vdv89,
ck90,
j90}.

The true evolutionary picture, however, may not be this simple.
Of the stars with IRAS colors suggesting a detached shell (i.e., a 
60 $\mu$m excess)
a significant fraction are semi-regulars, leading \citet{wj88}
to suggest that semi-regulars are transition objects, associated
with thermal pulses and dredge-up.
However, 
\citet{ie95}
found that contamination by cirrus
may be responsible for most of the observed 60 $\mu$m excesses,
negating this argument.
It is true, however, that some semi-regulars do have very extended
envelopes.
In their study of circumstellar envelopes resolved in the IRAS
database, \citet{ypk93b} found that most of the resolved
stars are semi-regulars, while most of the unresolved stars
are Miras.  They conclude that the most likely explanation is
that the semi-regulars have been losing mass longer. 
This suggests that semi-regulars are not always in an earlier
evolutionary stage from Miras.

This idea is supported by theoretical modeling of the 
$[$12$]$ $-$ $[$25$]$ to K $-$ $[$12$]$ 
colors and IRAS spectra of semi-regulars and Miras by \citet{ik00} and \citet{mik01}.
Their work indicates that semi-regulars lack hot dust in their
envelopes, implying a decrease in the mass-loss rate in the last
$\sim$100 years.
Furthermore, very long term light curves of a few variable stars show 
apparent mode switching, with 
dramatic changes in their light curves from
long-period, high amplitude variations to short-period lower amplitude variations
\citep{bedding98}.
Thus the evolutionary connection between semi-regulars and Miras is still uncertain.

\subsection{OH/IR Stars}

Another complication to the evolutionary scenario outlined above
is the fact that it is
difficult
to distinguish between mass sequences and evolutionary
sequences.
As noted above, 13 stars in our sample are likely massive stars not on the AGB
($>$6$-$8 M$_{\sun}$ main sequence mass), based on their classification as 
non-carbon luminosity class I or II stars or variability
type SRc or Lc stars.
For OH/IR stars, another mass indicator
is the velocity separation $\Delta$V of the 1612 MHz lines,
which is a measure of the expansion velocity of the circumstellar shell.
This velocity separation 
is correlated with galactic latitude and random motion relative
to galactic rotation, and
is 
therefore correlated with main sequence mass, with stars
with the highest $\Delta$V being mainly 
M supergiants
\citep{b81}.
In Figure 21, 
for the
OH/IR stars in our sample with 
$\Delta$V values available from \citet{c01},
we plot $\Delta$V vs. the DIRBE flux ratios
for the six shortest DIRBE wavelengths.
The star with the largest $\Delta$V in our sample by far is VY CMa,
the sole known OH/IR supergiant in the sample, 
with $\Delta$V = 65~km~s$^{-1}$ \citep{c01}.
VY CMa is quite variable in the optical, varying between magnitude 6.5 and 9.6
at V (GCVS). It is also somewhat variable in the DIRBE database (see Figures 4I $-$ J),
but less so than the Miras,
varying by only 0.16 $\pm$ 0.02 magnitudes
at 4.9 $\mu$m,  0.10 $\pm$ 0.01 magnitudes at 12 $\mu$m,
and 0.11 $\pm$ 0.01 magnitudes at 25 $\mu$m.
After VY CMa, there is a large gap in $\Delta$V, with the
next largest $\Delta$V belonging to the Mira variable WX Psc.

At all wavelengths, a slight correlation is seen in Figure 21, in that
the stars become redder as $\Delta$V increases.  
Such a correlation with
F$_{12}$/F$_{25}$ 
has been seen before 
(e.g., \citet{let90, c01});
the DIRBE data show that it is also present at the other infrared 
wavelengths as well,
particularly F$_{4.9}$/F$_{12}$.
The existence of an OH
maser requires a large dust column density to shield the OH molecules from
photodissociation \citep{hg82}.
As discussed by \citet{let90},
when the expansion velocity is large, dust shielding for the OH molecules
is diluted, reducing the OH column density and so the 
probability of an observable OH maser.
Thus for large expansion velocities, only objects with high dust column densities
(and therefore redder colors) are seen as OH masers.
The bluest OH masers are therefore those with smaller $\Delta$V.

Note that, in addition to VY CMa, which
stands alone, there appears to be two distinct groups of OH/IR stars
present in these plots: those with $\Delta$V $\le$ 15~km~s$^{-1}$
and those between 20 and 40~km~s$^{-1}$, with a gap 
between the
two groups.  The colors of the second group, on average, are
clearly redder than those in the first group.
The fundamental difference between these two groups may be
main sequence mass;
\citet{bh83}
conclude that OH/IR stars
with smaller values of $\Delta$V 
($\le$15~km~s$^{-1}$)
have lower main sequence masses 
than those with $\Delta$V between 20 and 40~km~s$^{-1}$.

Another parameter that may be related to main sequence mass is the
IRAS spectral type.
\citet{vk88}
found that stars with silicate absorption features
tend to lie in the Galactic Plane, while silicate emission stars are found
at both high and low galactic latitude.
In their study of OH/IR stars, \citet{c01}
found 
that the LRS type `A' (absorption) stars mainly lie in the 
Galactic plane, in contrast to the type `E' (emission) stars.
These studies concluded that LRS type `A' stars
are more massive than many of the type E stars, and that
the LRS types may be a mass sequence,
rather than an evolutionary sequence.  Within the high mass stars,
there may be evolution from LRS type E to A, but lower mass
stars do not reach the A stage.
This conclusion is reinforced by
the study of \citet{hc01}, who
found that the type E OH/IR stars appear to have
a different luminosity-period relation than the type A OH/IR stars.

The sample studied in the current paper, which is selected
to be at relatively high latitude ($|$b$|$ $>$ 5$^{\circ}$), 
only contains one LRS type A source, 
the unspecified variable
OH~338.1+6.4, an M star at b = 6.4$^{\circ}$.
This source has a much more moderate $\Delta$V than
VY CMa, 26~km~s$^{-1}$, similar to many of the type E Miras.
It is much more variable in the DIRBE database than VY CMa,
varying by about 0.6 magnitudes at 12 $\mu$m and 0.2 magnitudes
at 25 $\mu$m, typical of a Mira.
Compared to the LRS type `E' stars with similar $\Delta$Vs (and
therefore presumably similar main sequence masses), OH~338.1+6.4
is redder in the DIRBE F$_{4.9}$/F$_{12}$ and F$_{12}$/F$_{25}$
ratios
(Figure 21), suggesting that it may be in a later stage
of evolution than the type E stars.  Thus this may
be evidence for evolution from type E to type A stars
within the higher main sequence mass range.

The other two OH/IR stars without a variability class
(OH~329.8-15.8 and OH~348.2-19.7) were identified as OH/IR stars
by 
\citet{g89},
who noted that they
are optically invisible and 
variable in the IRAS database. 
In the \citet{c01}
database, 
their values of 
$\Delta$V are also moderate, 29~km~s$^{-1}$ and 24~km~s$^{-1}$, respectively.
In the DIRBE database, OH~348.2-19.7 is strongly variable, by
1.11 $\pm$ 0.08 magnitude at 4.9 $\mu$m and 1.20 $\pm$ 0.16  
magnitudes at 12 $\mu$m (Figures 4K and L).
In contrast, OH~329.8-15.8 is only marginally variable in the DIRBE
database (0.31 $\pm$ 0.07 magnitude at 4.9 $\mu$m and 0.25 $\pm$ 0.08 magnitudes
at 12 $\mu$m; see Figures 4M and N).

Recent work by \citet{l02} has called into question
the idea that OH/IR
stars are always very evolved stars in the last stage of evolution
before becoming planetary nebulae.
His follow-up study of 328 
stars previously detected in the 1612 MHz
OH line yielded four non-detections, implying
that those four stars are now `dead' OH/IR stars: the
1612 MHz line has disappeared.
Since these stars still have other characteristics
of AGB stars, this implies that stars pass through the
OH/IR phase more than once.  A thermal pulse may trigger
a short period of heavy mass loss, sufficient to allow
temporary OH maser activity.  In this scenario,
relatively blue F$_{12}$/F$_{25}$ colors are expected
\citep{l02}.

Our sample contains two OH/IR stars, 
U Men and R Crt, which are classified
as semi-regulars rather than Miras.
They
have relatively low values of $\Delta$V (8 and 20~km~s$^{-1}$,
respectively; \citet{c01}).
Their infrared colors
similar to those of other semi-regulars (see Table 6) and are
also similar to OH/IR Miras with similarly-low values of
$\Delta$V (Figure 21).
The redder of these two stars,
the SRa star U Men, has 
a Mira-like V amplitude of 2.9 and a relatively
long period of 407 days (GCVS), and $\Delta$V of $\sim$
20~km~s$^{-1}$ \citep{c01},
in the gap between
the low mass and higher mass stars.
It also varies strongly in the infrared (see Figures 3L and M).
It also varies strongly in the infrared (see Figures 3L and M).
The bluer star, the SRb star R Crt, has a V amplitude
of only 1.4 and a much shorter period of 160 days (GCVS),
and has a small $\Delta$V of 8~km~s$^{-1}$ \citep{c01}.
R Crt is included in the \citet{c01}
catalog based on the possible
detection by \citet{dbb73}
of a weak 2 Jy 1612 MHz OH line.
However,
\citet{e01}
were not able to detect the 1612 MHz
line to a limit of 0.2 Jy, although the 
1667 and 1665 MHz lines are strong.
If the initial detection was valid,
this may be another example of a `dead' OH/IR star.

Based on studies of OH variability in
very dusty OH/IR stars,
it has been concluded 
that OH/IR stars become non-variable
just before the planetary nebula stage \citep{o84,
b87, vdvh88}.
As discussed above,
of our four reddest OH/IR stars, at least one is likely not
an AGB star.
Of the reminding three, 
two are clearly strongly variable in the DIRBE database.
We note, however, that our sample is biased
towards stars bright in the mid-infrared, thus may be somewhat
prejudiced against extremely evolved objects.
\citet{o84}
find a turn-off of variability
at about 2.5~log(F$_{12}$/F$_{25}$) of $-$0.8.  In our sample, only 
the LRS type A star
OH
338.1 +6.4 is redder than this criteria; this star
is strongly variable in the DIRBE database.

\subsection{Carbon-Rich Stars }

A similar infrared color-variability type trend is visible within the carbon-rich stars,
in that 
carbon-rich Miras are clearly 
redder than carbon SRb stars.  
For carbon stars, increasing optical depth of the
circumstellar shell, and therefore
increasing redness, is expected as the star evolves
\citep{ck90}.
Carbon-rich Lb stars have infrared colors
consistent with those of carbon-rich SRb stars.  
This agreement has been noted before \citep{klh96}.
There are too few carbon-rich SRa
stars in this sample to make a conclusive statement about their infrared
colors.

The carbon stars in our sample without previously known variability 
have colors similar to those of known Miras, suggesting that
many of these objects may also be Miras, but are too optically faint to
have been monitored and classified.
This is confirmed by inspection of the DIRBE light curves
of these objects (for example, see Figures 4G and H).  Of the thirteen carbon-rich stars without variability
information, twelve are clearly variable in the DIRBE light curves,
and the thirteenth is probably variable.
The DIRBE amplitudes of variation
for these objects are large,
on average 1 magnitude at 4.9 $\mu$m,
0.6 magnitudes at 12 $\mu$m, and 0.5 magnitudes at 25 $\mu$m.
These are slightly larger than
the amplitudes of variation of the known Miras in the sample (Table 7).
The three carbon stars with unspecified variability also have large amplitudes
(Table 7).
This implies that
essentially all of the very red carbon stars in the IRAS
catalog are likely previously unclassified Mira variables.
For 7 of these 13 carbon-rich stars, we are able
to make a rough lower limit to the pulsation period from the DIRBE
database.
These lower limits range from 360 to 500 days, showing, as expected,
that these very red stars have long pulsation periods.

\subsection{Carbon-Rich vs. Oxygen-Rich Stars }

Table 6 shows that, for F$_{1.25}$/F$_{2.2}$ and F$_{2.2}$/F$_{3.5}$, 
carbon-rich stars of a particular variability type
are noticeably redder than oxygen-rich stars of the same variability type.
The same is true for F$_{3.5}$/F$_{4.9}$, with the exception of the SRbs,
where the oxygen-rich stars are redder.
At F$_{12}$/F$_{25}$, for all variability types the oxygen-rich
stars are redder, while at F$_{4.9}$/F$_{12}$ 
there is little difference
in the colors of the two chemical groups.
These are consistent with the results of 
\citet{g93},
who found that oxygen-rich stars are redder than carbon-rich
stars in $[$12$]$ $-$ $[$25$]$, but
bluer in K $-$ L.   
As discussed by \citet{wk98}, redder colors
for carbon stars relative to oxygen-rich stars are expected, 
due to the different
opacities of silicate and carbon grains.  Carbon dust is more opaque
at short wavelengths than silicate dust, causing carbon-rich stars
to be more obscured and have higher infrared excesses for the same 
mass loss rates.
At the shortest wavelengths,
the redder F$_{1.25}$/F$_{2.2}$ colors of carbon stars can be
accounted for by different molecular opacities in the stellar atmospheres
\citep{mgc03}.

In the DIRBE dataset, the three non-supergiant
OH/IR stars in our sample without known variability types
are as red or redder than the carbon-rich Miras, with both
groups 
having large inferred mass loss rates.  
The large mass loss rates
of these two groups
were discussed by \citet{ck90},
who concluded that
both extreme carbon stars and very red OH/IR stars are 
immediate planetary nebulae progenitors.
This implies that there are two different evolutionary tracks
for AGB stars, one for carbon-rich stars and one for oxygen-rich
stars.
According to current theories of stellar evolution,
when stars first reach the AGB, they are all oxygen-rich; later,
some become carbon-rich due to dredge-up \citep{ir83, ck90}.
Only stars with main sequence masses between $\sim$1.5 $-$ 4 M$_{\sun}$
become carbon stars;
higher and lower mass stars do not
go through this phase
\citep{gvj95, ml02}.
For comparison, the typical OH/IR star has a main sequence mass
of $\sim$1 M$_{\sun}$, according to \citet{likkel89}.

\subsection{S Stars and Stars with Mixed Chemistry}

There are only four stars in our sample with optical type S,
three of which are Miras.
The Mira S stars in our sample have DIRBE colors more like the
oxygen-rich Miras rather the
carbon-rich stars.  
It has been suggested \citep{dj89}
that S stars are transition
objects, in the process of converting from oxygen-rich to
carbon-rich.  This idea is supported by the AGB evolutionary
calculations of 
\citet{gvj95}.
The S Miras in our sample tend to have somewhat
larger amplitudes
of variations compared to oxygen-rich and carbon-rich
Miras (Table 7), but this result
may not be statistically significant,
since the number of S stars is very small.

There has been quite a bit of discussion in the literature
about carbon stars with oxygen-rich circumstellar shells
(i.e., \citet{lm86, wj86,
dj89}).
These stars may be transition objects, stars
that have recently become carbon stars and still have relic
oxygen-rich shells.
Our sample does not contain any carbon-rich stars with
oxygen-rich shells, however, it contains something less expected:
an oxygen-rich star with a carbon-rich shell.
This star, GY Cam, is classified as M6 by \citet{lgb47} and
as SR: in the GCVS.
GY Cam is quite variable in the DIRBE data,
with an observed amplitude of 0.96 $\pm$ 0.04
magnitudes at 4.9 $\mu$m and 0.68 $\pm$ 0.16 magnitudes
at 12 $\mu$m,
consistent with the Miras in the sample.

This star is not unique;
a few other M stars with SiC-rich shells were identified
by \citet{sgw90}.
If these IRAS identifications and optical types are correct,
then these stars present an intriguing puzzle. 
As the carbon stage follows the oxygen-rich stage,
it is difficult to explain how a carbon-rich shell can form around
an oxygen-rich star.  
One possibility is that they are binary
or confused systems, and the shell did not originate from
the observed optical star.  
Another possibility suggested by \citet{sgw90}
is that these stars have C/O ratios $\sim$ 1 (i.e., they are
mis-classified S stars), which can have carbon-rich shells.
Follow-up observations are needed
to check the IRAS association and optical spectral type of GY Cam.

\subsection{Massive Stars}

The massive stars in the sample generally show less
variation in the DIRBE database than the AGB
stars (see Figures 13 $-$ 18 and Table 7).
The most variable massive star in the sample is 
the M3/M4II OH/IR star VY CMa, which,
as discussed earlier, varies by 0.16 magnitudes at 4.9 $\mu$m,
$\sim$0.1 magnitude at
12 and 25 $\mu$m, and $\sim$ 3 magnitudes at V. 

All six of the SRc stars in the sample show evidence
of $>$3$\sigma$ variability in the DIRBE database,
with observed variations of $\sim$0.05 $-$ 0.1 magnitude
at 1.25 $-$ 4.9 $\mu$m.  The highest S/N light curves
in this class are those of $\alpha$~Ori, which varies
by 0.12 $\pm$ 0.01 magnitudes at 4.9 $\mu$m
and 0.10 $\pm$ 0.01 magnitude at 12 $\mu$m (Figures 4O and P).
The two Lc stars and the two bright giants classified
as SRa or SRb are less clearly variable in the DIRBE
data ($\sim$3$\sigma$).
The two bright giants/supergiants not previously
identified as variable ($\alpha$ Car and c Pup)
are not variable in the DIRBE data.

\section{Summary}

We have used
the COBE database to extract
near- and mid-infrared 
flux densities for 207 of
the brightest infrared point sources
in the sky, significantly expanding our earlier
pilot study of 38 stars.
The majority of these objects are asymptotic giant
branch stars.  From the DIRBE light curves, 
we made estimates of the infrared variability of these
stars.
At 4.9 $\mu$m, 138 of the 207 sources are observed to be
variable at the 3$\sigma$ level, and 106 sources at the 
5$\sigma$ level.
In this study,
we have identified 18 variable stars not previously
known to be variable.

The DIRBE colors become increasingly redder along
the sequence optically-bright slightly-varying giant $-$$>$ SRb $-$$>$
SRa $-$$>$ Mira $-$$>$ OH/IR stars,
with carbon stars
being redder than oxygen-rich stars for the same variability type.
The colors and variability of the Lb stars are consistent
with those of the SRb stars, supporting previous assertions
that these are the same type of object.
The DIRBE variability also increases along this sequence;
optically-bright giants have little to no observed variability
in the DIRBE data, while the reddest stars vary by up to 
$\sim$1 magnitude at 4.9 $\mu$m and $\sim$0.8 magnitudes at 12 $\mu$m.

Table 1 and 5
are available in electronic form,
from the Astronomical Journal electronic edition
at
http://www.journals.uchicago.edu/AJ/ or at
http://www.etsu.edu/physics/bsmith/dirbe/.

We thank the COBE team
for making this project possible.  We are especially
grateful to Nils Odegard, for developing the software used to
extract the DIRBE light curves.  
We also thank Danny Camper, an 
undergraduate at East Tennessee State University, for
help with the data acquisition.
We are grateful to Steve Price, Dave Leisawitz, Mark Giroux,
and Don Luttermoser for helpful suggestions.
This research has made use of the SIMBAD database,
operated at the CDS, Strasbourg, France, as well
as the Astronomical Data Center at NASA Goddard Space
Flight Center, the NASA Astrophysics Data System
at the Harvard-Smithsonian Center for Astrophysics, 
the VizieR service at the CDS, Strasbourg, France,
and the Infrared Science Archive, operated by
the Jet Propulsion Laboratory, California Institute
of Technology.
We have also made use of the electronic
versions of the General Catalogue
of Variable Stars, the New Suspected Variable Stars 
Catalogue, and the Supplement to this catalogue, provided
by the Sternberg Astronomical Institute at Moscow
State University.
This research was funded by
National Science Foundation POWRE
grant AST-0073853 and NASA LTSA grant NAG5-13079.

\vfill
\eject

%% Generally speaking, only the figure captions, and not the figures
%% themselves, are included in electronic manuscript submissions.
%% Use \figcaption to format your figure captions. They should begin on a
%% new page.

\clearpage

%% No more than seven \figcaption commands are allowed per page,
%% so if you have more than seven captions, insert a \clearpage

%% There must be a \figcaption command for each legend. Key the text of the
%% legend and the optional \label in curly braces. If you wish, you may
%% include the name of the corresponding figure file in square brackets.
%% The label is for identification purposes only. It will not insert the
%% figures themselves into the document.
%% If you want to include your art in the paper, use \plotone.
%% Refer to the on-line documentation for details.

\includegraphics{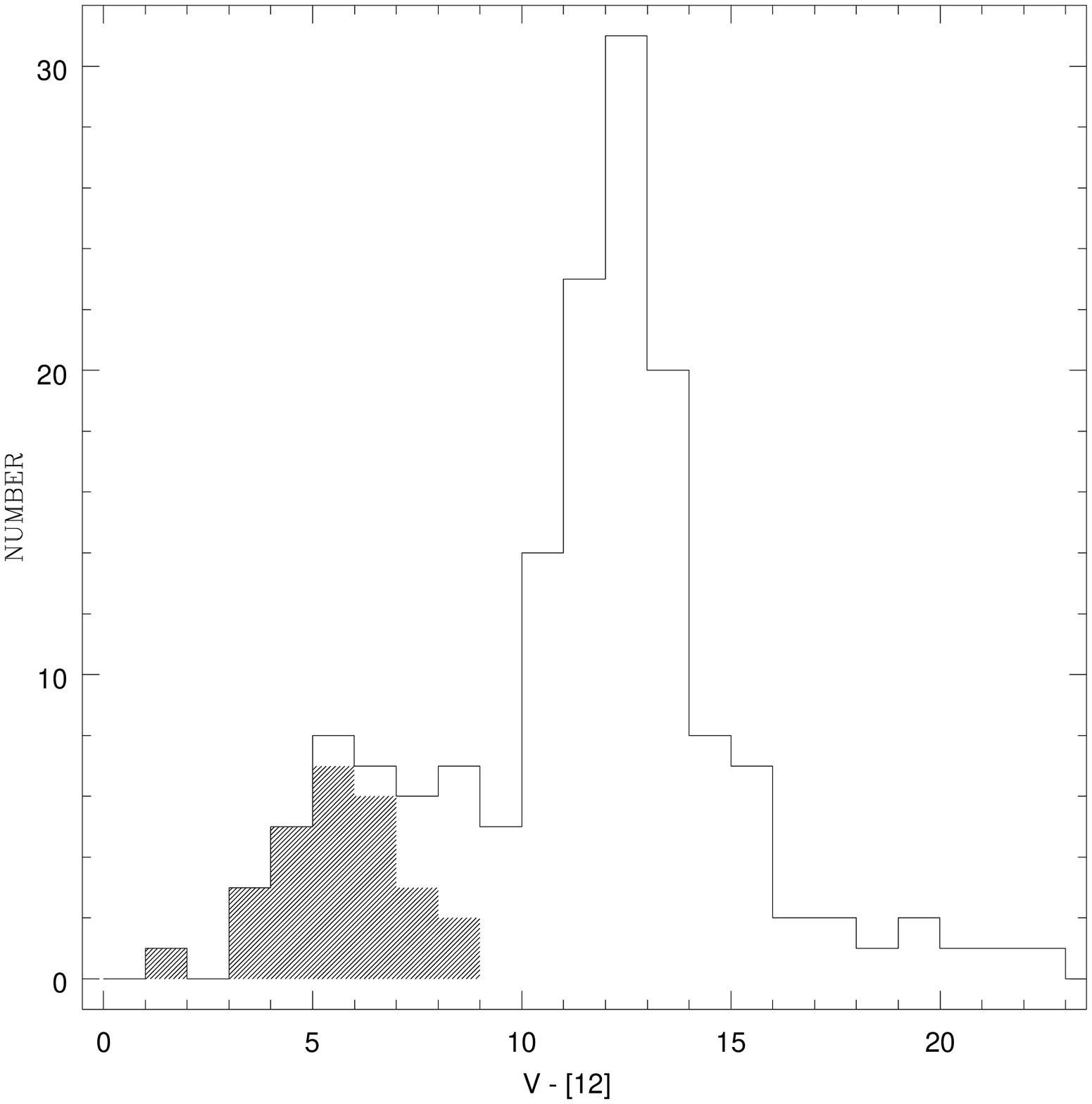}
\figcaption{
The distribution of 
V $-$ $[$12$]$ 
for the 155 sources in our sample with V magnitudes available.
The 12 $\mu$m magnitudes were calculated from the IRAS
Point Source Catalogue 12 $\mu$m flux densities assuming 
zero magnitude corresponds to 28.3 Jy (Beichman et al. 1988).
The shaded region marks the 27 stars with IRAS LRS spectral
type of `S' (dominated by photospheric emission).
The bluest star is the F0 II star $\alpha$ Car (Canopus).
Note that this histogram is likely incomplete at
V $-$ $[$12$]$ $>$ 11.8, which corresponds to a V = 10 magnitude
star at our 12 $\mu$m flux limit.
}

\includegraphics{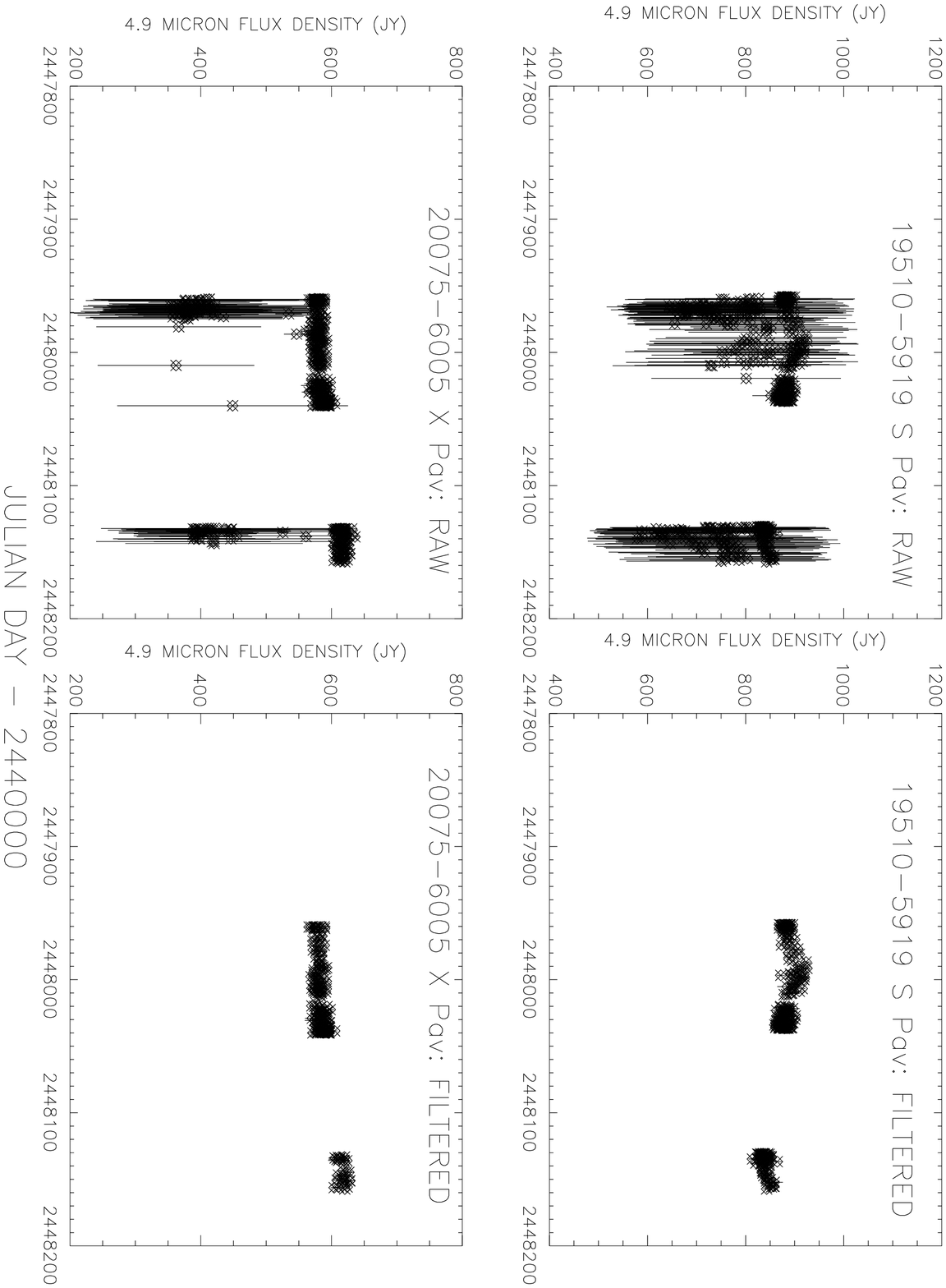}
\figcaption{
Examples of raw vs. filtered 4.9 $\mu$m DIRBE light curves 
for a few selected
objects, the SRa star S Pav
and the SRb star X Pav.  The filtering process is described in the text.
}

\figcaption{
Some representative filtered light curves.  
See text for more details.
}

\figcaption{
Some more representative filtered light curves.  
See text for more details.
}

\includegraphics{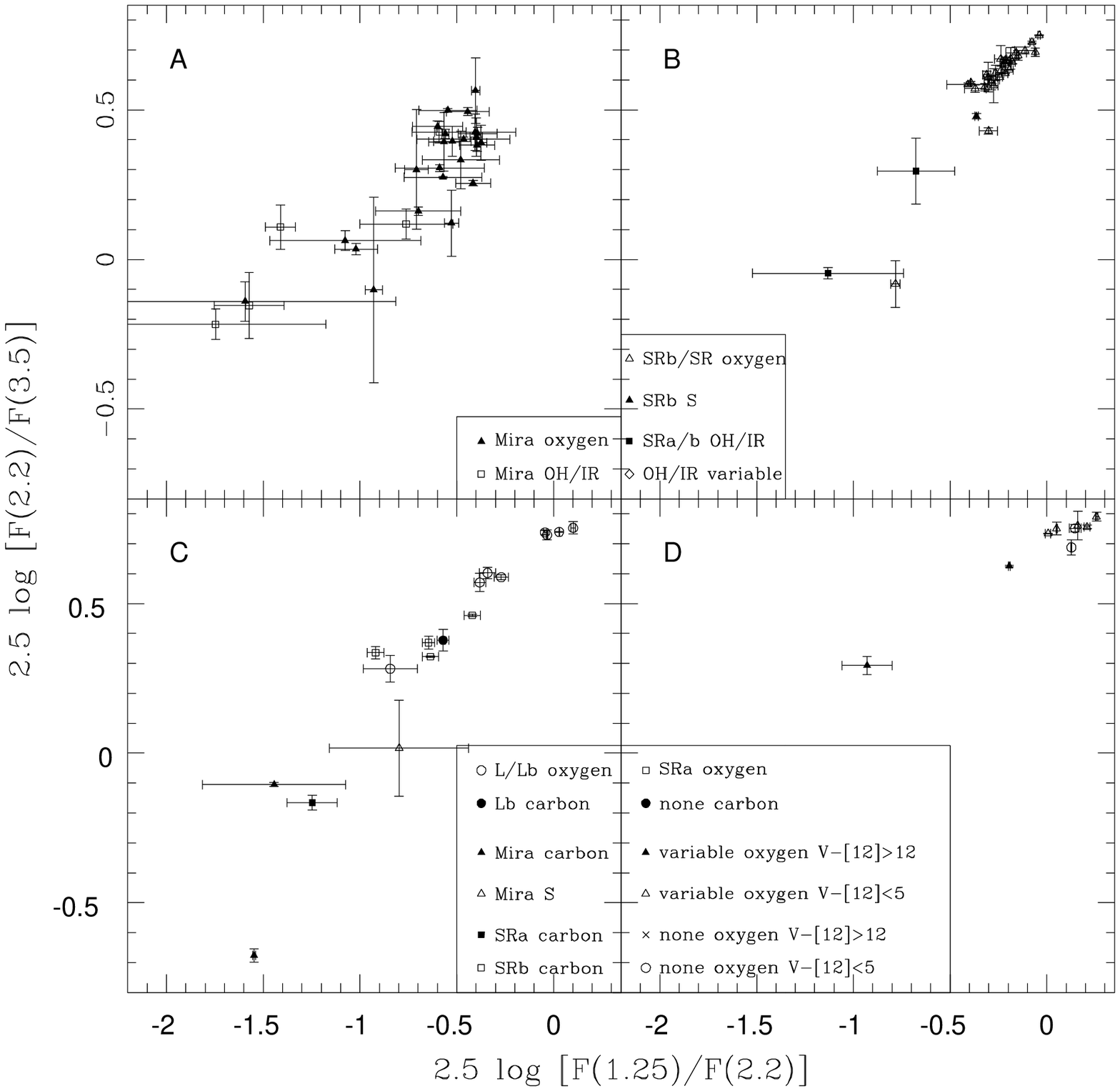}
\figcaption{
The time-averaged DIRBE 2.2 $\mu$m/3.5 $\mu$m vs.
1.25 $\mu$m/2.2 $\mu$m color-color plot, for types
of stars associated with the AGB.  
For clarity,
four panels with different types of objects are shown.
The symbols corresponding to different types of stars are 
as shown 
(`none' refers to stars not previously identified as variable.)
The errorbars
shown are the standard deviations of the observed infrared colors of the
stars,
and therefore include both intrinsic color variations
and measurement uncertainties.
This plot only includes stars with average S/N $>$ 5 at all three
wavelengths.
}

\includegraphics{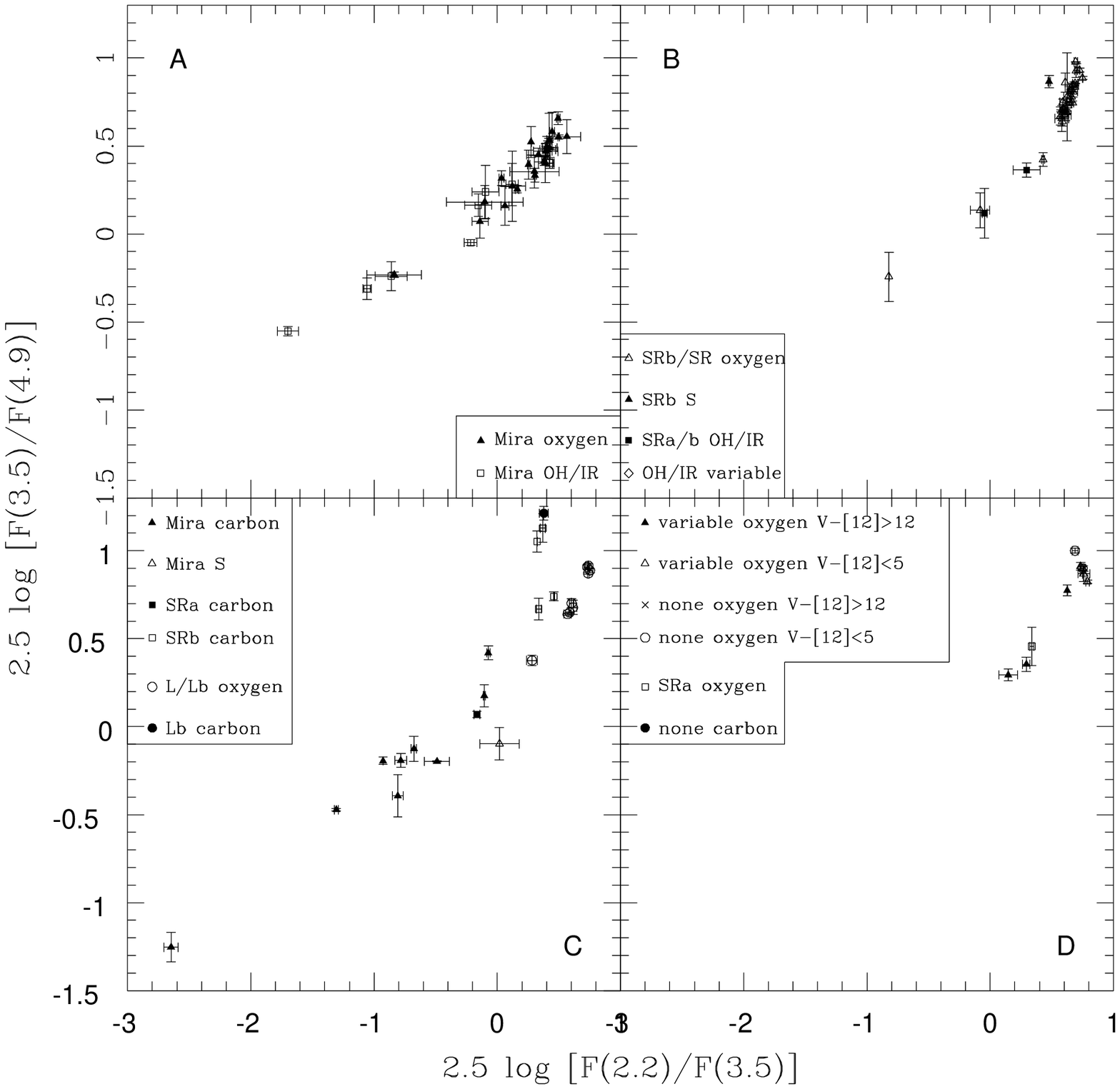}
\figcaption{
The time-averaged DIRBE 3.5 $\mu$m/4.9 $\mu$m vs.
2.2 $\mu$m/3.5 $\mu$m color-color plot.  
The symbols are as shown.
The errorbars
shown are the standard deviations of the observed infrared colors 
of the stars, and therefore include both intrinsic color variations
and measurement uncertainties.
This plot only includes stars with average S/N $>$ 5 at all three
wavelengths.
}

\includegraphics{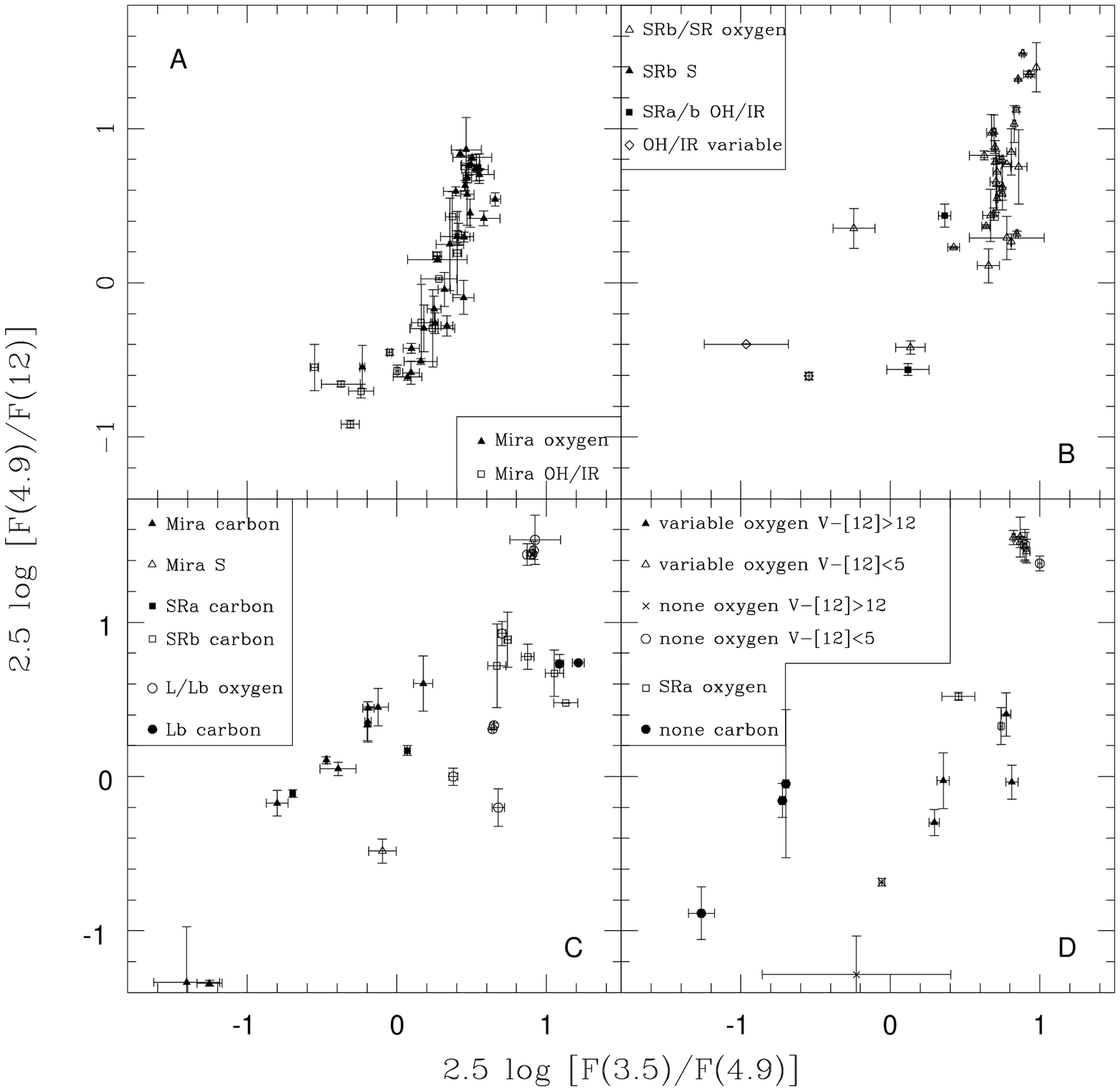}
\figcaption{
The time-averaged DIRBE 4.9 $\mu$m/12 $\mu$m vs.
3.5 $\mu$m/4.9 $\mu$m color-color plot.  
The symbols are as shown.
The errorbars
shown are the standard deviations of the individual flux
measurements, and therefore include both intrinsic variations
of the sources as well as measurement uncertainties.
This plot only includes stars with average S/N $>$ 5 at all three
wavelengths.
}

\includegraphics{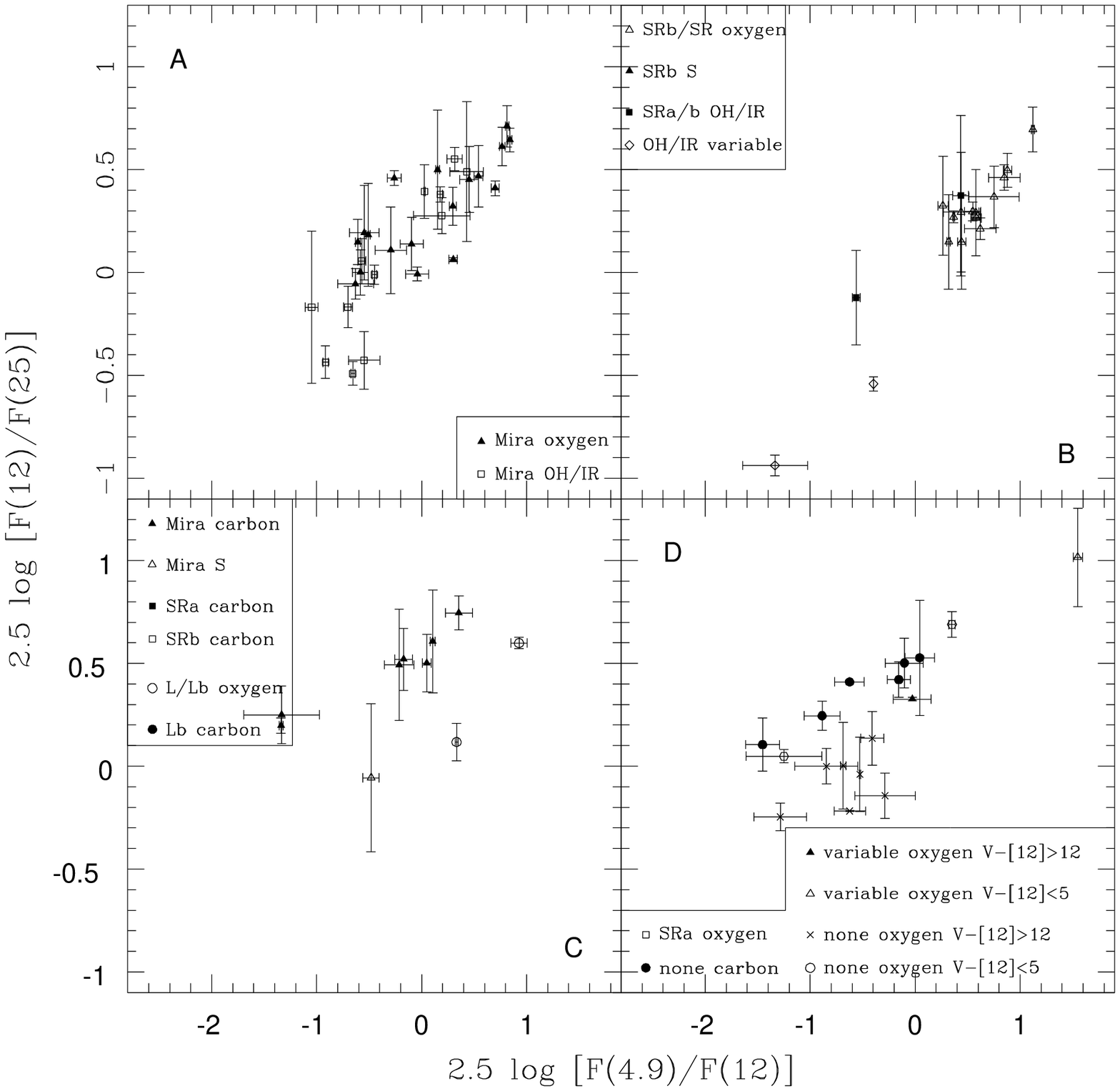}
\figcaption{
The time-averaged DIRBE 12 $\mu$m/25 $\mu$m vs.
4.9 $\mu$m/12 $\mu$m color-color plot.  
The symbols are as shown.
The errorbars
shown are the standard deviations of the infrared colors
and therefore include both intrinsic color variations
and measurement uncertainties.
}

\includegraphics{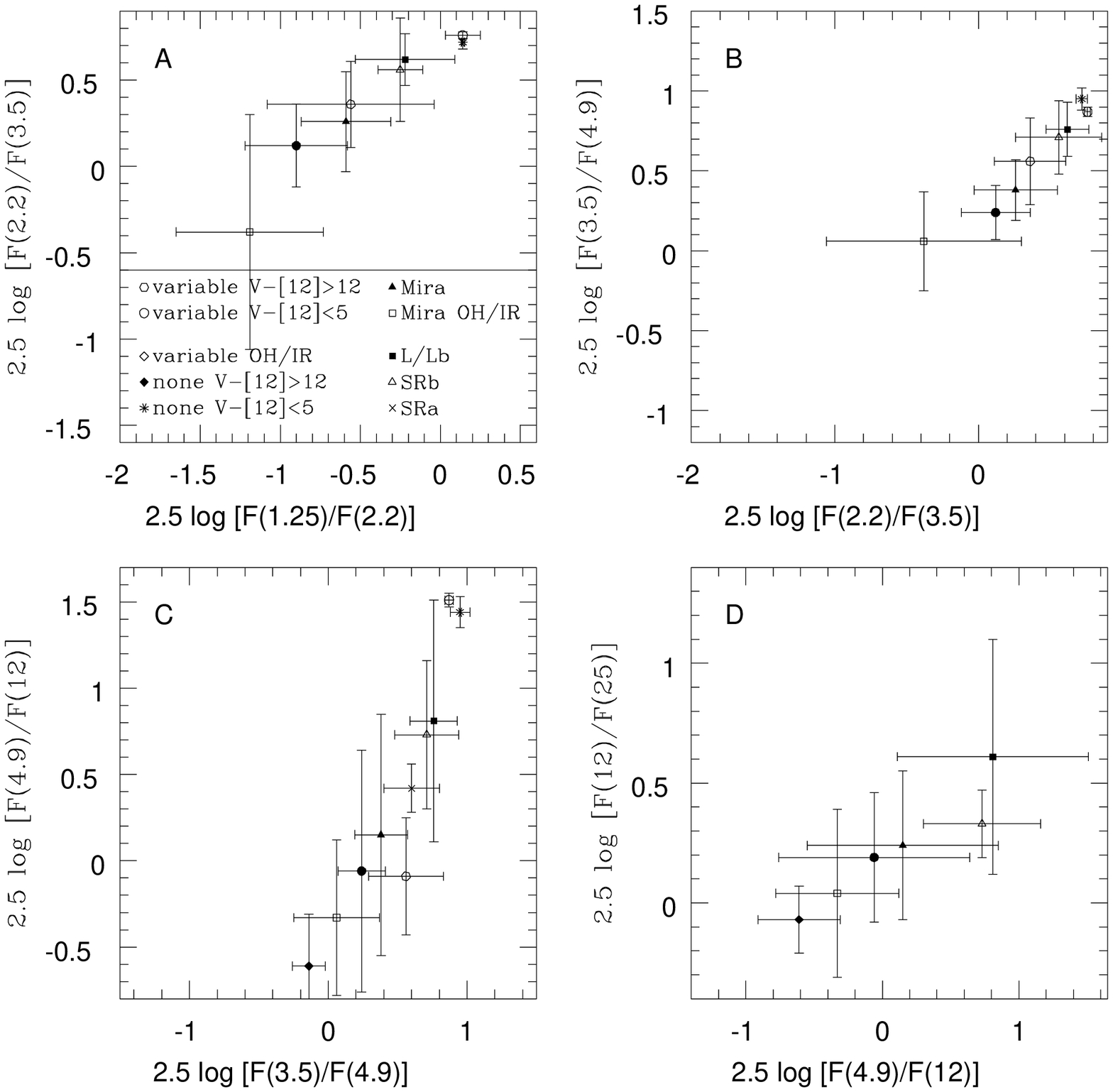}
\figcaption{
This Figure shows the means of the time-averaged DIRBE colors for different
types of oxygen-rich stars associated with the AGB, along with
the rms spread (from Table 6).
The symbols identified in panel A correspond to datapoints in all
four panels. 
Categories with less than two high quality measurements
at all three wavelengths are not shown.
}

\includegraphics{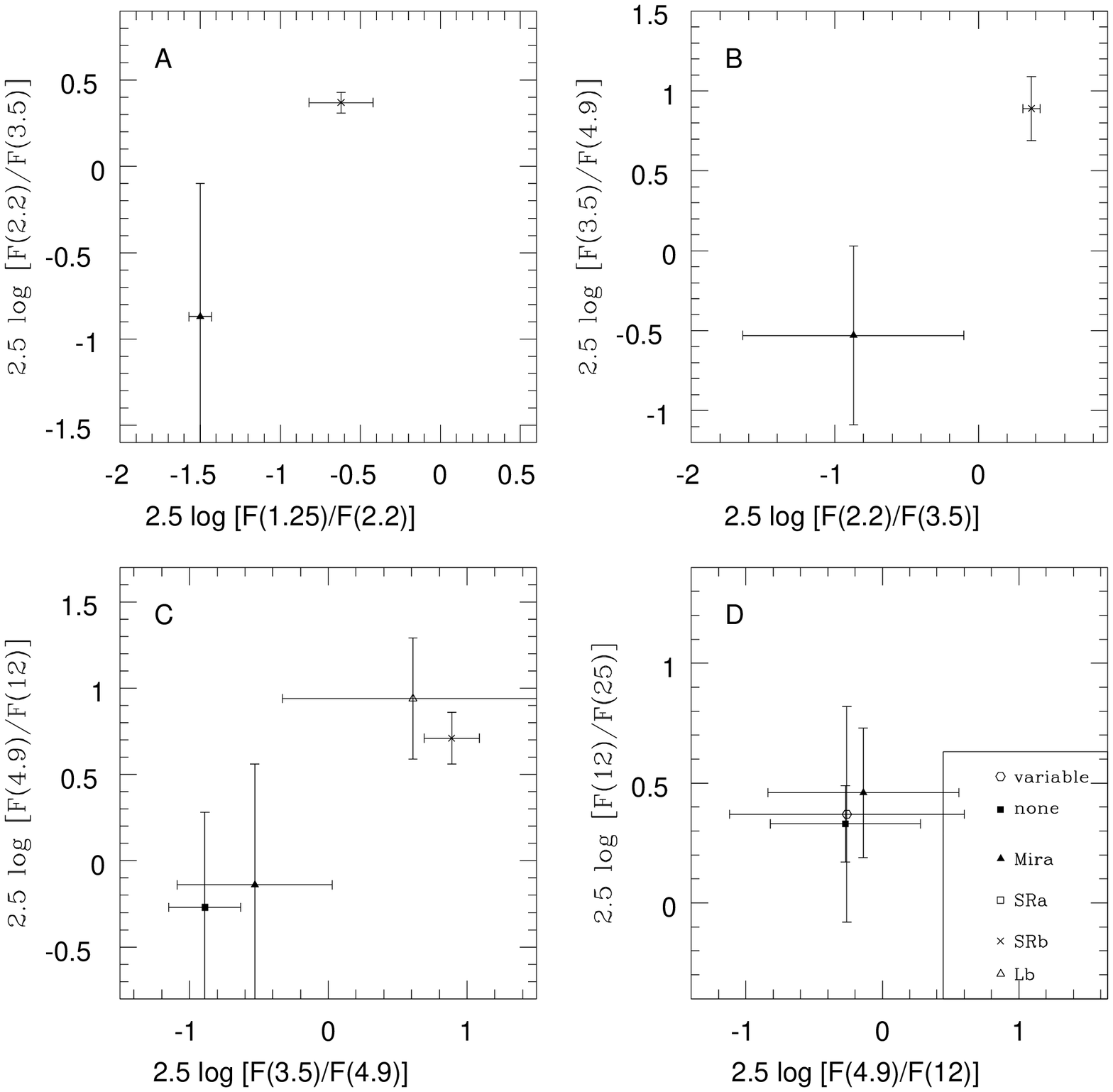}
\figcaption{
This Figure shows the mean of the time-averaged DIRBE colors for the different
types of carbon-rich stars in the sample, along with
the rms spread (from Table 6).
The symbols identified in panel D correspond to datapoints in all
four panels. 
Categories with less than two high quality measurements
at all three wavelengths are not shown.
In Panel B, two datapoints are superposed (the means for the unspecified
variables and the stars not previously known to be variable are the same).
}

\includegraphics{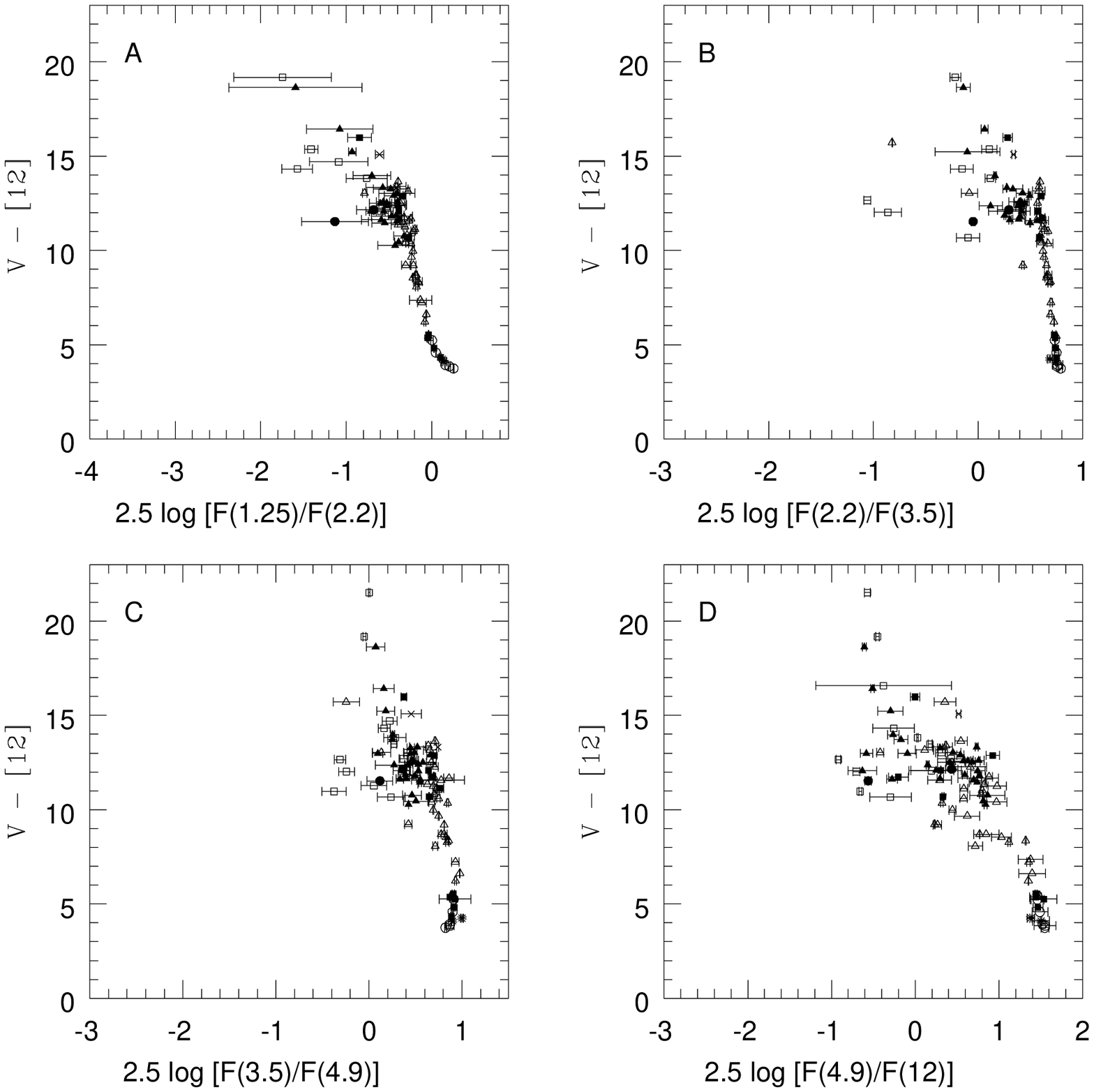}
\figcaption{
This Figure shows the mean DIRBE colors for the different
types of oxygen-rich stars associated with the AGB, plotted against
V $-$ $[$12$]$.
The symbols are as in Figure 9.
}

\includegraphics{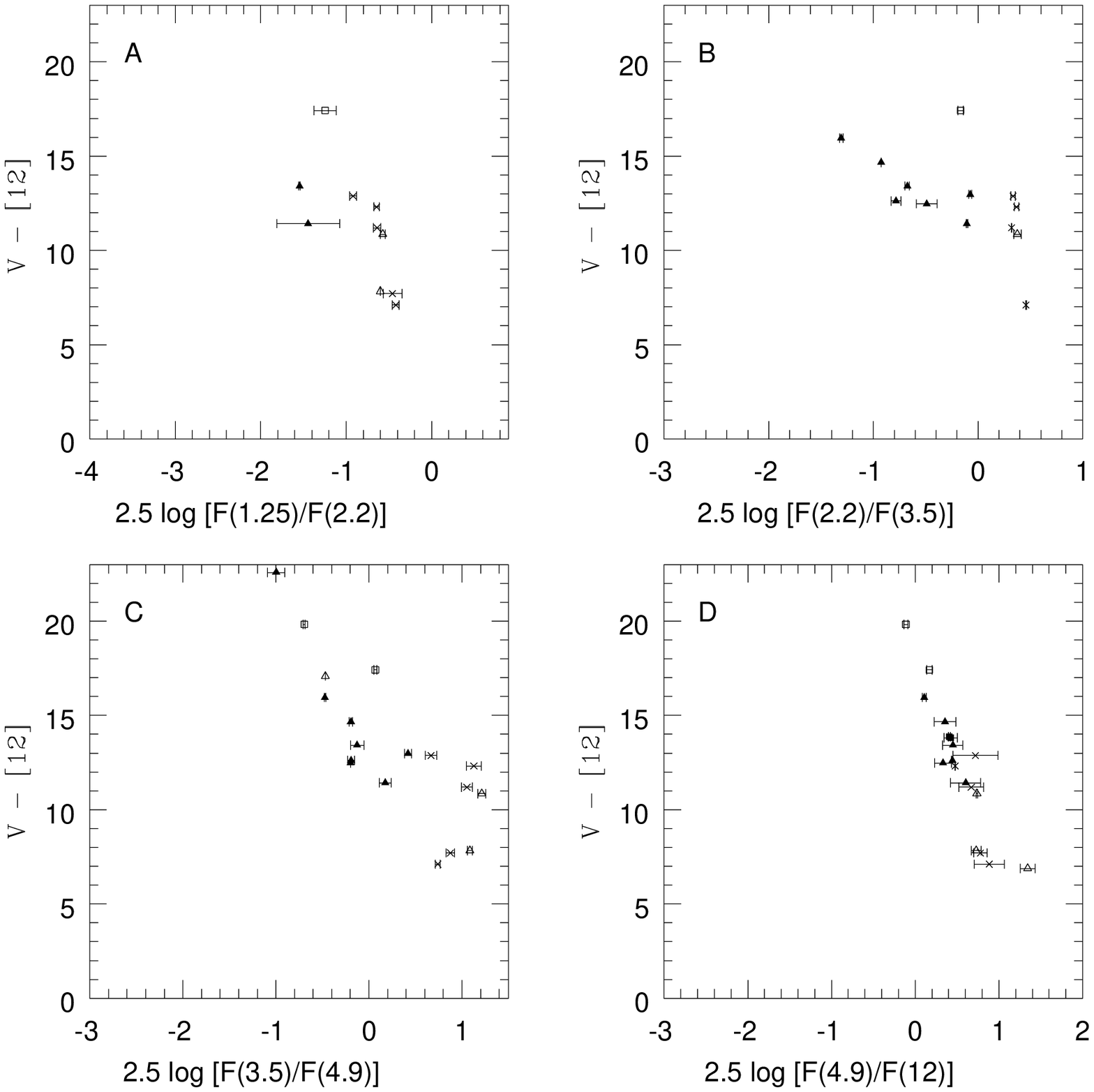}
\figcaption{
This Figure shows the mean DIRBE colors for the different
types of carbon-rich stars associated with the AGB, plotted against
V $-$ $[$12$]$.
The symbols are as in Figure 10.
The SRa (open square) at 
V $-$ $[$12$]$ = 19.8 is RW LMi;
the SRa at 
V $-$ $[$12$]$ = 19.8 is V Hya.
The very red Mira (filled triangle) at
V $-$ $[$12$]$ = 22.6 is LP And.
}

\includegraphics{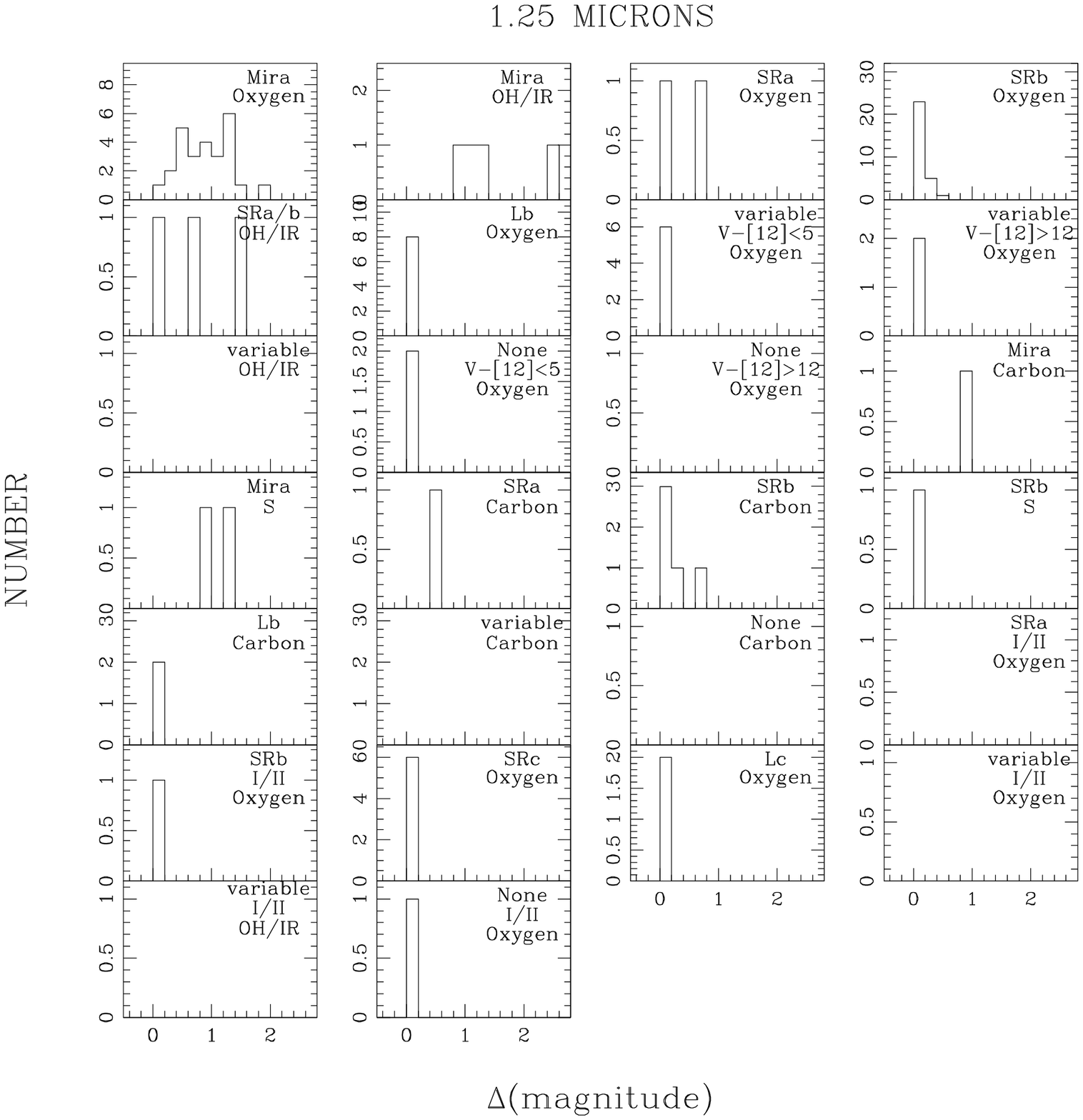}
\figcaption{
Histograms of the {\it observed} 1.25 $\mu$m amplitudes of variation in the light
curves of various types of stars in our sample.
Note that these amplitudes are not necessarily the full amplitude of variation
for the stars, since the DIRBE data may not cover the full pulsation period.
These plots only include stars with at least 5$\sigma$ DIRBE flux densities
at 1.25 $\mu$m at minimum light.
If a panel does not have a histogram, it means no star in that group
was detected above 5$\sigma$ at 1.25 $\mu$m.
The average uncertainty in the observed $\Delta$(mag) at 1.25 $\mu$m for the sample
is 0.06 magnitudes.
}

\includegraphics{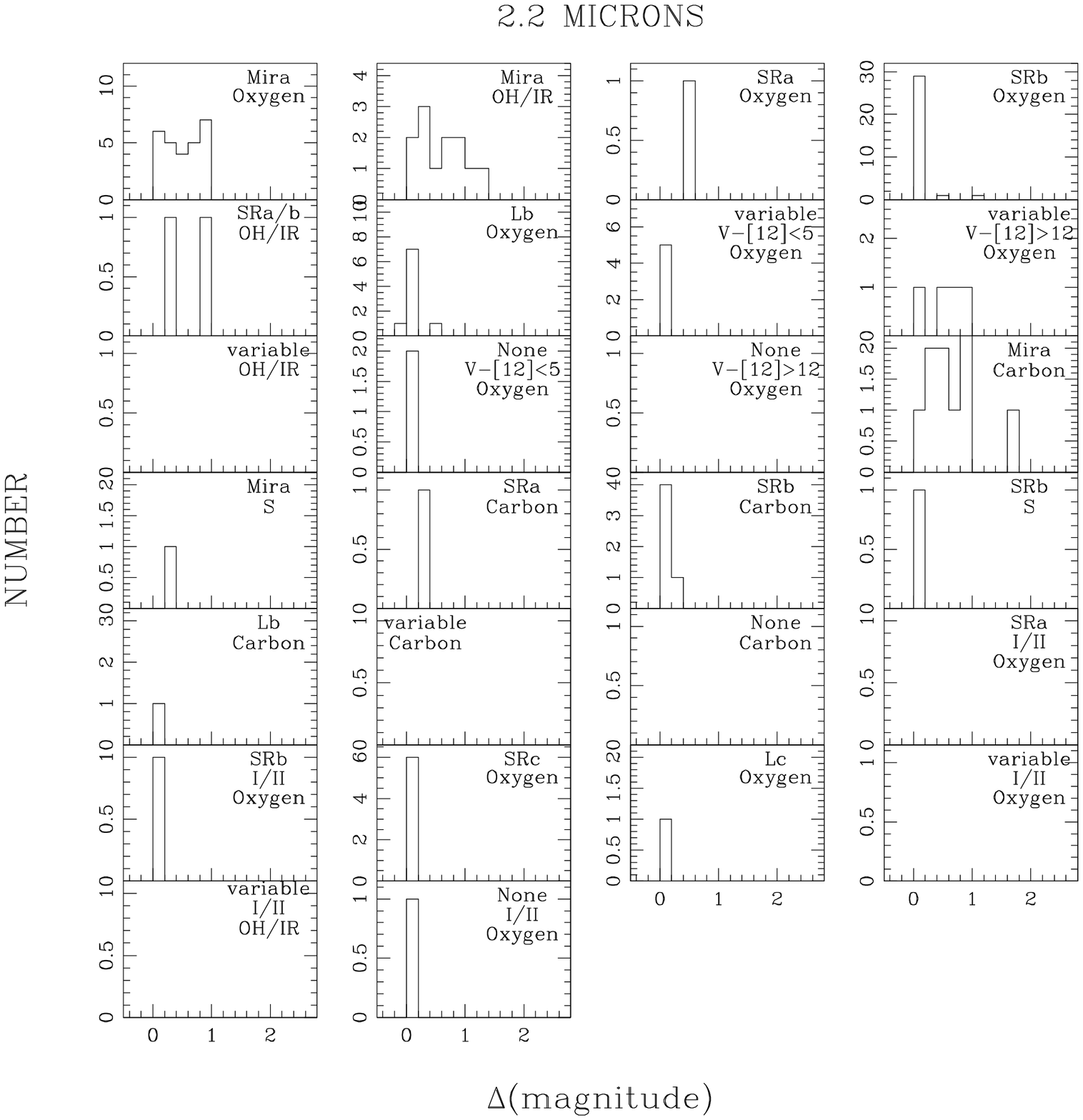}
\figcaption{
Histograms of the {\it observed} 2.2 $\mu$m amplitudes of variation in the light
curves of various types of stars in our sample.
Note that these amplitudes are not necessarily the full amplitude of variation
for the stars, since the DIRBE data may not cover the full pulsation period.
These plots only include stars with at least 5$\sigma$ DIRBE flux densities
at 2.2 $\mu$m at minimum light.
If a panel does not have a histogram, it means no star in that group
was detected above 5$\sigma$ at 2.2 $\mu$m.
The average uncertainty in the observed $\Delta$(mag) at 2.2 $\mu$m for the sample
is 0.05 magnitudes.
}

\includegraphics{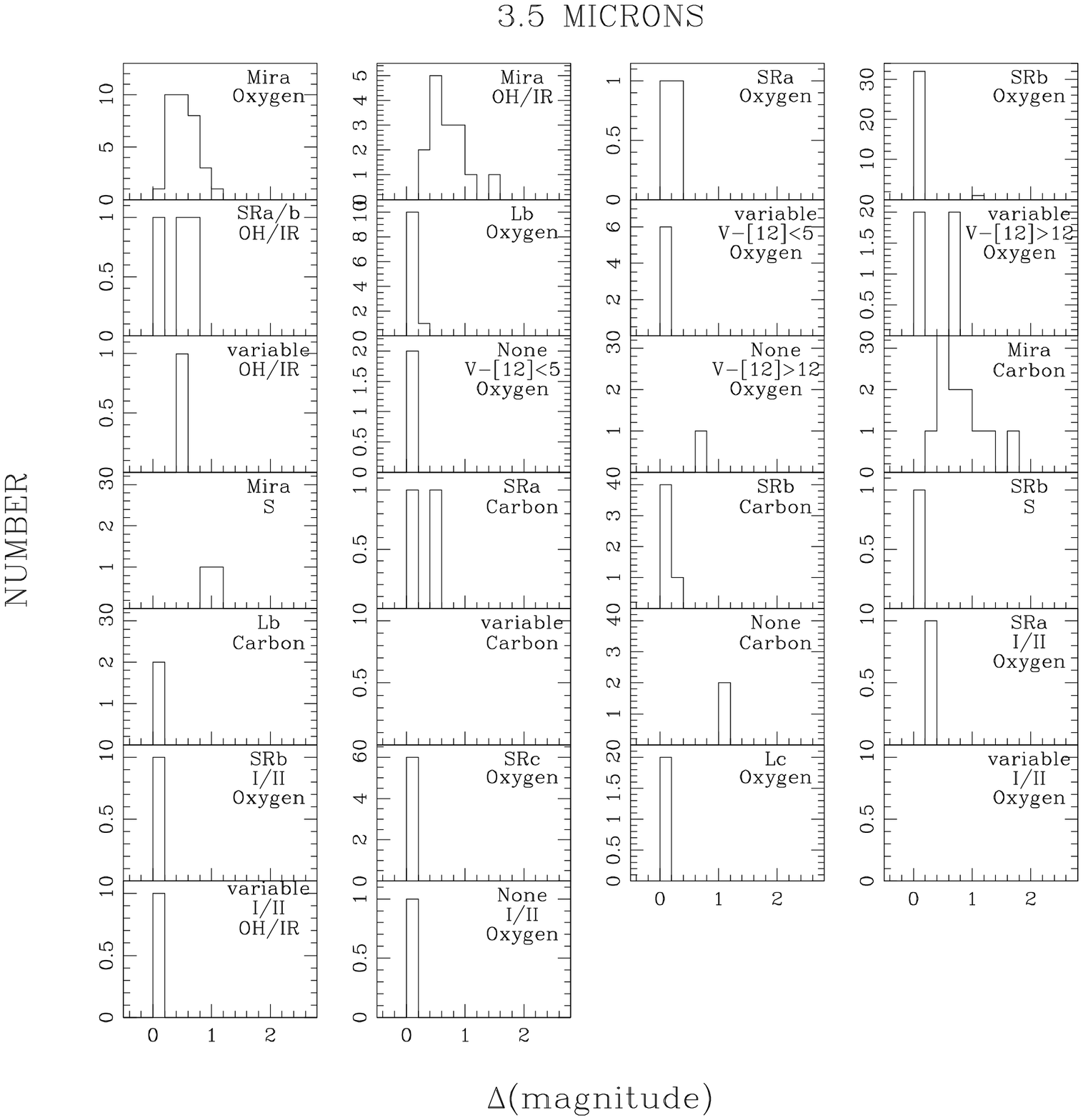}
\figcaption{
Histograms of the {\it observed} 3.5 $\mu$m amplitudes of variation in the light
curves of various types of stars in our sample.
Note that these amplitudes are not necessarily the full amplitude of variation
for the stars, since the DIRBE data may not cover the full pulsation period.
These plots only include stars with at least 5$\sigma$ DIRBE flux densities
at 3.5 $\mu$m at minimum light.
If a panel does not have a histogram, it means no star in that group
was detected above 5$\sigma$ at 3.5 $\mu$m.
The average uncertainty in the observed $\Delta$(mag) at 3.5 $\mu$m for the sample
is 0.06 magnitudes.
}

\includegraphics{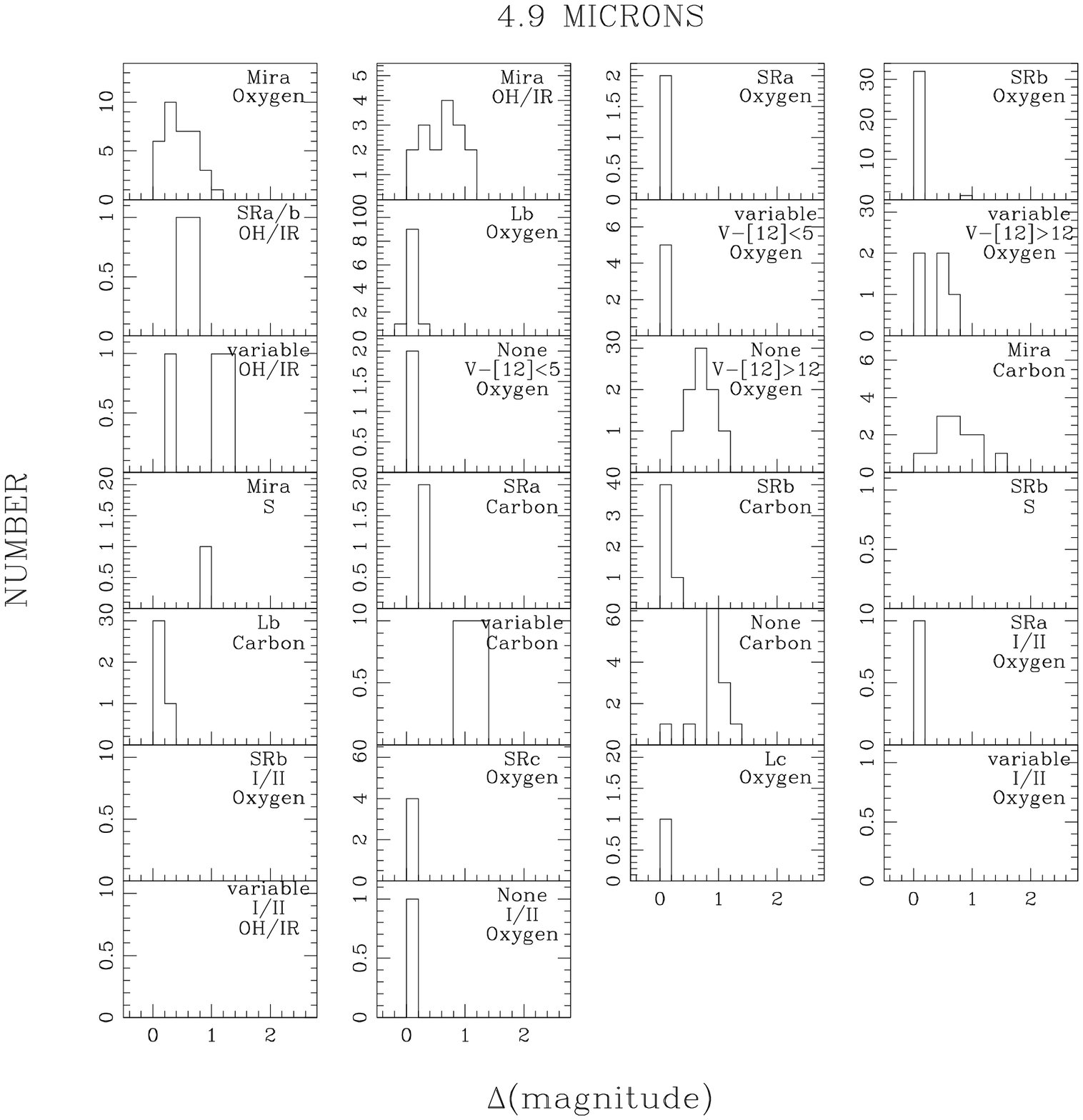}
\figcaption{
Histograms of the {\it observed} 4.9 $\mu$m amplitudes of variation in the light
curves of various types of stars in our sample.
Note that these amplitudes are not necessarily the full amplitude of variation
for the stars, since the DIRBE data may not cover the full pulsation period.
These plots only include stars with at least 5$\sigma$ DIRBE flux densities
at 4.9 $\mu$m at minimum light.
If a panel does not have a histogram, it means no star in that group
was detected above 5$\sigma$ at 4.9 $\mu$m.
The average uncertainty in the observed $\Delta$(mag) at 4.9 $\mu$m for the sample
is 0.05 magnitudes.
}

\includegraphics{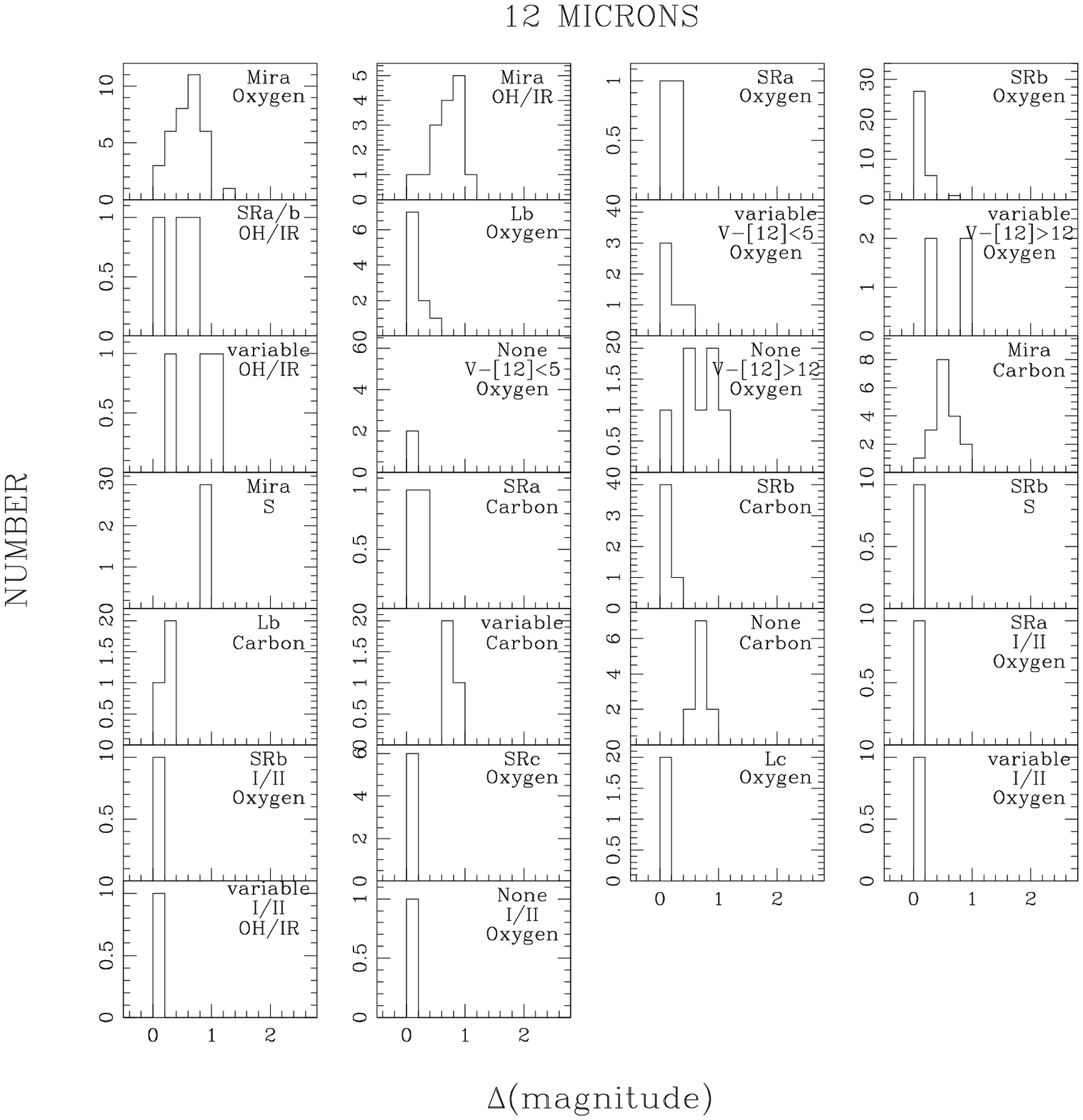}
\figcaption{
Histograms of the {\it observed} 12 $\mu$m amplitudes of variation in the light
curves of various types of stars in our sample.
Note that these amplitudes are not necessarily the full amplitude of variation
for the stars, since the DIRBE data may not cover the full pulsation period.
These plots only include stars with at least 5$\sigma$ DIRBE flux densities
at 12 $\mu$m at minimum light.
If a panel does not have a histogram, it means no star in that group
was detected above 5$\sigma$ at 12 $\mu$m.
The average uncertainty in the observed $\Delta$(mag) at 12 $\mu$m for the sample
is 0.11 magnitudes.
}

\includegraphics{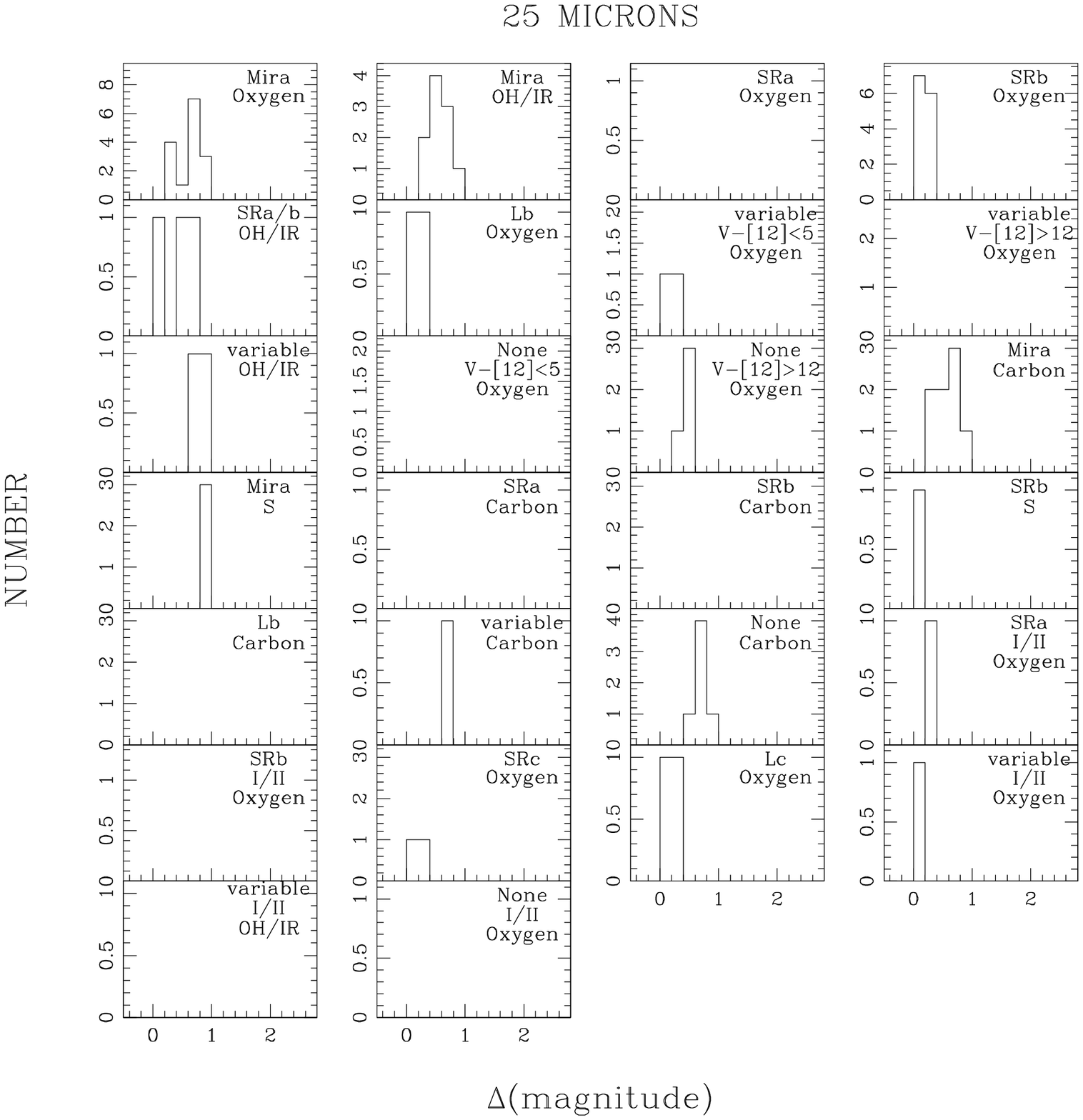}
\figcaption{
Histograms of the {\it observed} 25 $\mu$m amplitudes of variation in the light
curves of various types of stars in our sample.
Note that these amplitudes are not necessarily the full amplitude of variation
for the stars, since the DIRBE data may not cover the full pulsation period.
These plots only include stars with at least 5$\sigma$ DIRBE flux densities
at 25 $\mu$m at minimum light.
If a panel does not have a histogram, it means no star in that group
was detected above 5$\sigma$ at 25 $\mu$m.
The average uncertainty in the observed $\Delta$(mag) at 25 $\mu$m for the sample
is 0.14 magnitudes.
}

\includegraphics{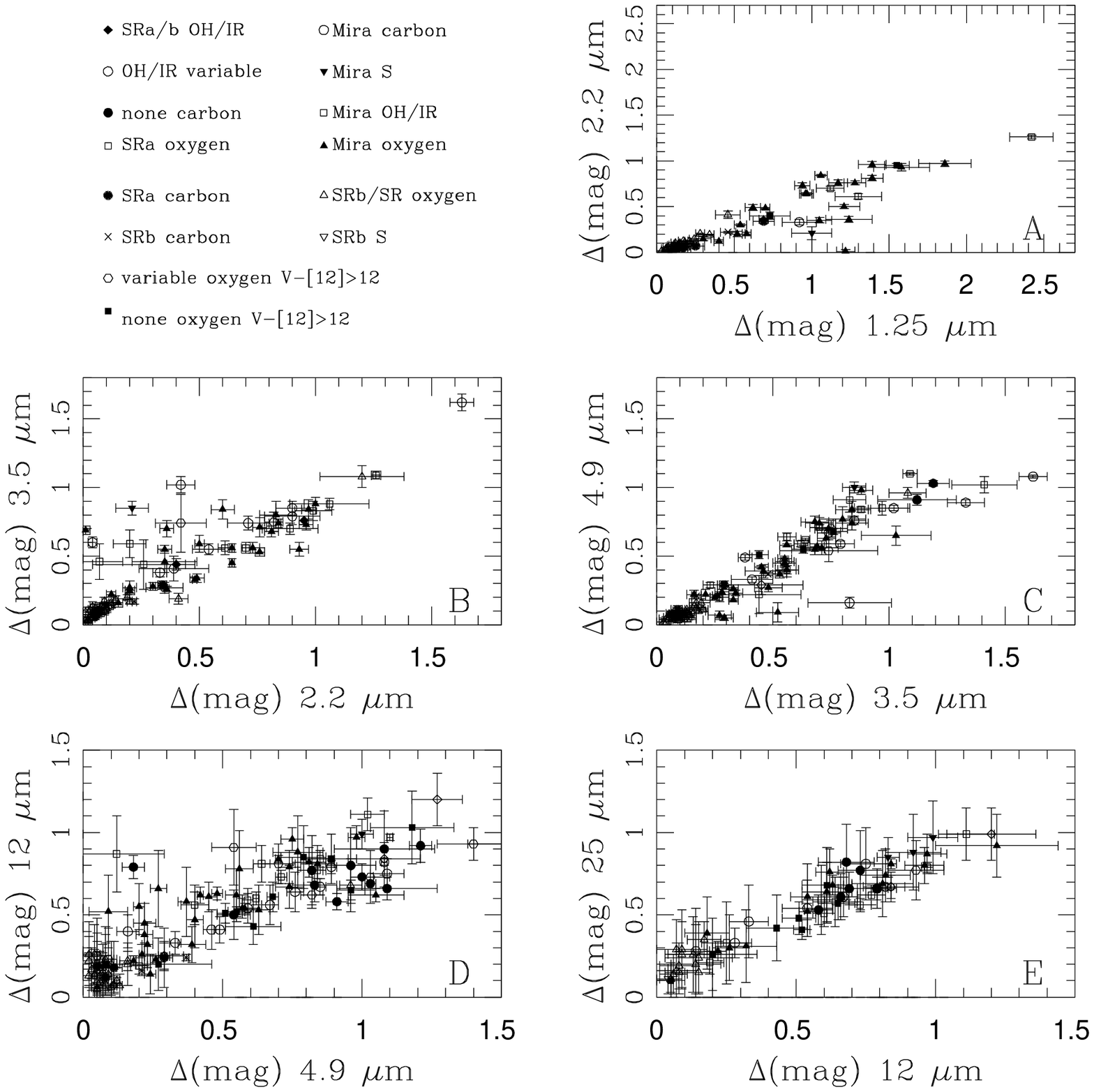}
\figcaption{
The {\it observed} amplitudes at the five shortest DIRBE wavelengths,
plotted against those at the next longest wavelength.
Only stars with greater than 5$\sigma$ flux densities at
their light curve minima are included.  
The symbols identified in the upper left
correspond to datapoints in all
five panels. 
}

\includegraphics{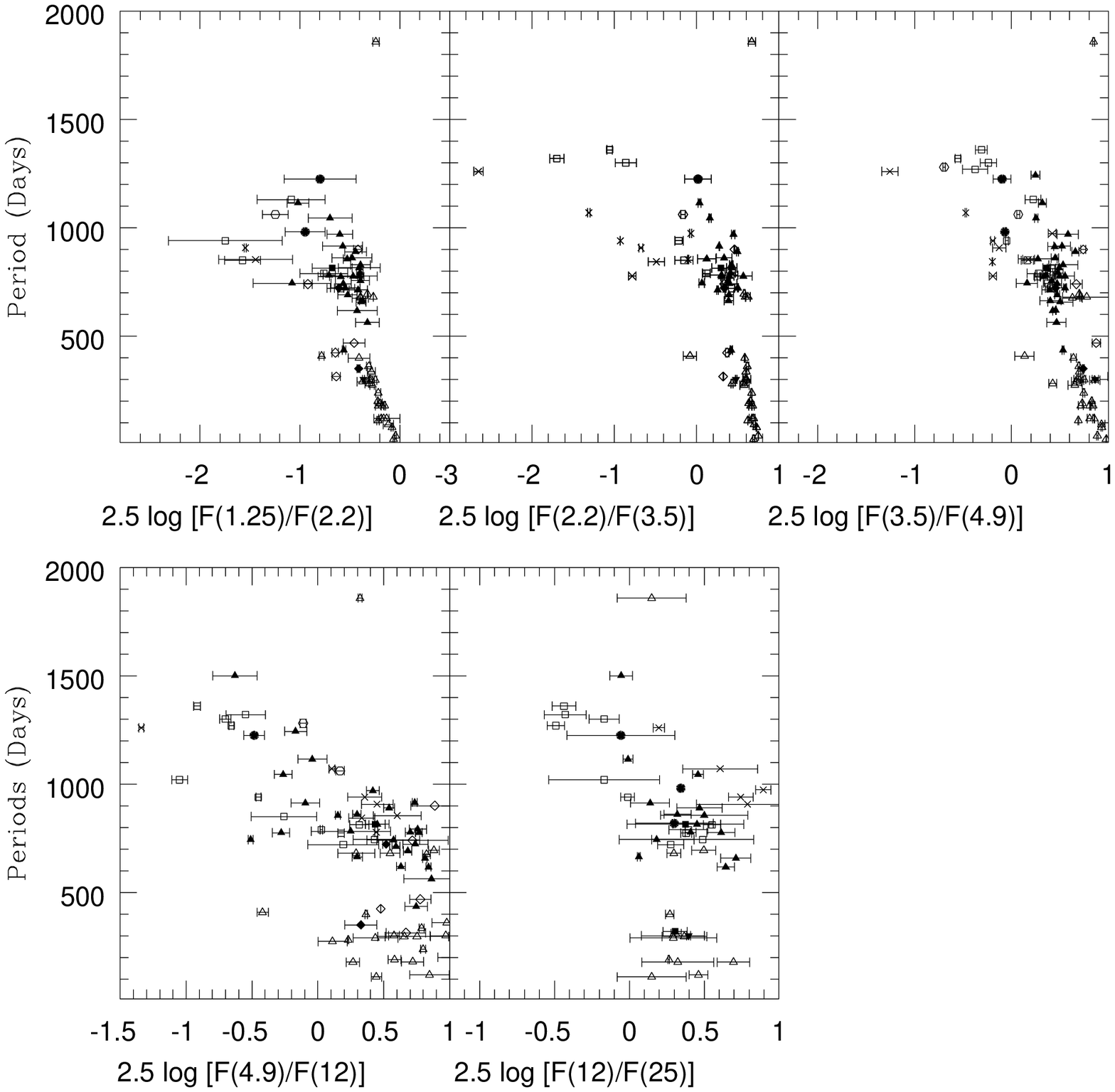}
\figcaption{
The relationship between the pulsation period
and the DIRBE colors, for the 103 likely AGB stars in the sample
with known periods.
The filled triangles are oxygen-rich Miras, the open squares
are Miras known to contain 1662 MHz OH masers, the crosses are
the carbon Miras, and the asterisks are the type S Miras.
Oxygen-rich SRb stars are open triangles, carbon-rich
SRa's are filled diamonds, SRa/b sources associated
with OH masers are filled squares, carbon-rich SRa's are open
hexagons, carbon-rich SRbs are open diamonds,
and SRb's that are optical type S stars are upside down
filled triangles.
}

\includegraphics{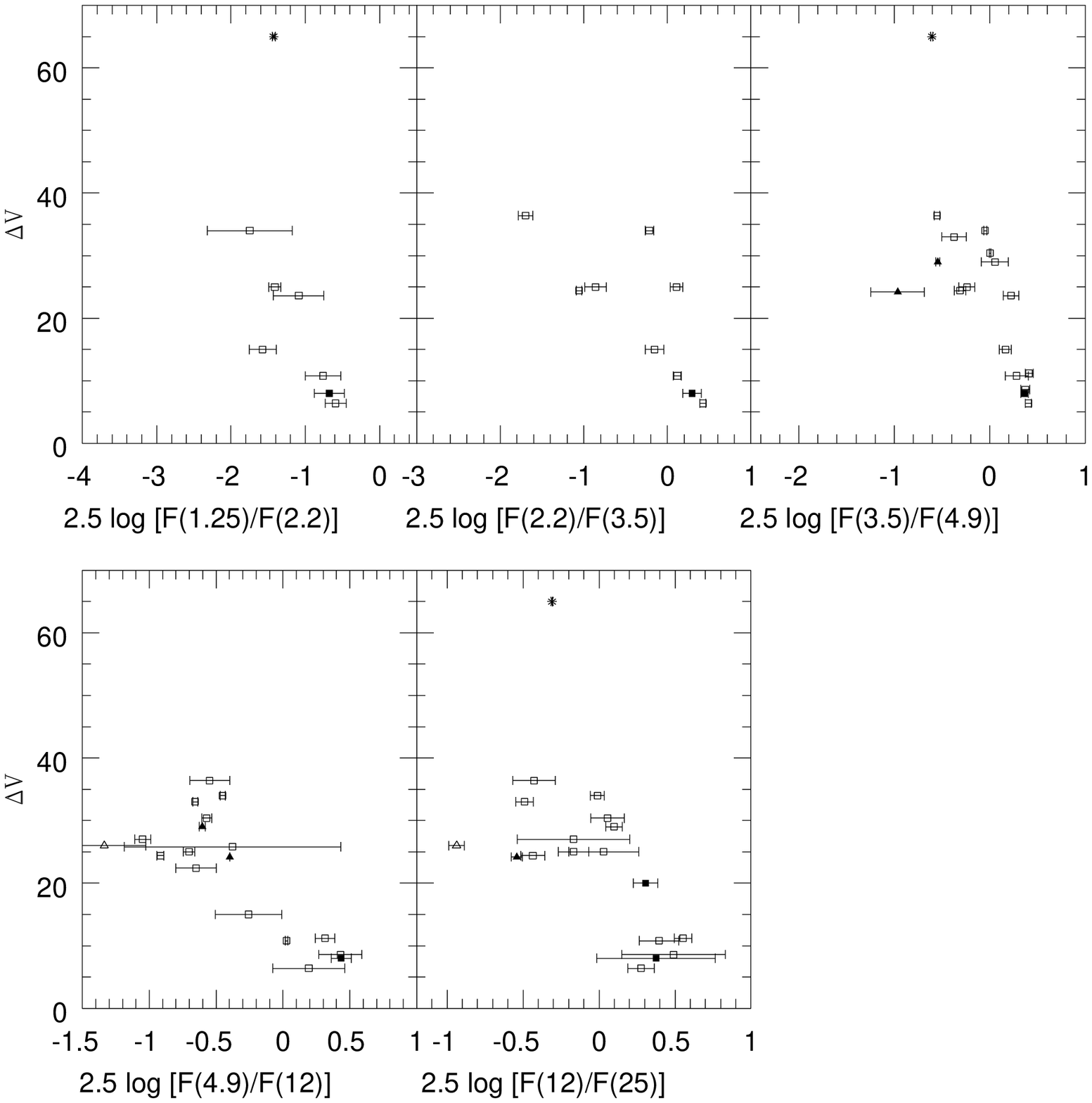}
\figcaption{
For the OH/IR stars with velocity separations $\Delta$V for the 1612 MHz
OH line available from Chen et al. (2001), 
$\Delta$V is plotted against the time-averaged DIRBE flux ratios for the six shortest
DIRBE wavelengths.
The Mira OH/IR stars are plotted as open squares and the semi-regular OH/IR
stars U Men and R Crt are filled squares. The unspecified
variable OH/IR stars are filled triangles, with the exception
of the sole LRS type `A' star in the sample, OH 338.1+6.4, which is
plotted as an open triangle.
VY CMa,
an unspecified variable OH/IR star known to be a supergiant,
is plotted as an asterisk.
}

%% Tables should be submitted one per page, so put a \clearpage before

%% Two options are available to the author for producing tables:  the
%% deluxetable environment provided by the AASTeX package or the LaTeX
%% table environment.  Use of deluxetable is preferred.
%%

%% Three table samples follow, two marked up in the deluxetable environment,
%% one marked up as a LaTeX table.

%% In this first example, note that the \tabletypesize{}
%% command has been used to reduce the font size of the table.
%% Note also that the \label command needs to be placed 

\clearpage

%% manuscript produces a one-column, double-spaced document:

%% preprint2 produces a double-column, single-spaced document:

\begin{deluxetable}{cccccccc}
\tabletypesize{\scriptsize}
\tablecaption{Sample\label{tbl-1}}
\tablewidth{0pt}
\tablehead{
\colhead{IRAS }&
\colhead{Other} & \colhead{Object}   & \colhead{Spectral}   &
\colhead{LRS}&
\colhead{V} &
\colhead{Period} 
\\
\colhead{Name}&
\colhead{Name} & \colhead{Type}   & \colhead{Type}   &
\colhead{Type}& 
\colhead{mag} &
\colhead{(Days)} 
\\
\hline
\hline
\multicolumn{7}{c}
{
Available in the Electronic Astronomical Journal at http://www.journals.uchicago.edu/AJ
}
}
\startdata
\enddata
\end{deluxetable}

\clearpage

%\documentclass[12pt,preprint]{aastex}

%% manuscript produces a one-column, double-spaced document:

%\documentclass[manuscript]{aastex}

%% preprint2 produces a double-column, single-spaced document:

%\documentclass[preprint2]{aastex}

%\begin{document}
 
\begin{deluxetable}{crrrrrrrrrrrrrr|r}
\tabletypesize{\scriptsize}
\tablecaption{VARIABILITY TYPES vs. OPTICAL SPECTRAL TYPES\label{tbl-2}}
\tablewidth{0pt}
\tablehead{
\colhead{} &
\multicolumn{14}{c}{\bf VARIABILITY TYPE}\\
\colhead{\bf OPTICAL}&
\colhead{\bf Mira} & \colhead{\bf SRa}   & \colhead{\bf SRb}   &
\colhead{\bf SRc} &
\colhead{\bf SR} &
\colhead{\bf Lb} &
\colhead{\bf Lc} &
\colhead{\bf L} &
\colhead{\bf variable} &
\colhead{\bf Z+S} &
\colhead{\bf IN} &
\colhead{\bf R} &
\colhead{\bf E} &
\colhead{\bf none} &
\colhead{\bf TOTAL} \\
\colhead{\bf TYPE}&
\\
}
\startdata
{\bf M }       & 58&5&33&6&3&10&2& & 9& 1& & & & 3 & 130\\      
{\bf S }       &  3& & 1& & &  & & &  &  & & & &   &   4\\     
{\bf C }       & 14&2& 5& & & 4& & & 1&  & & & & 4 &  30\\     
{\bf K }       &  1& &  & & & 1& & & 4&  & & &1& 1 &   8\\
{\bf F }       &   & &  & & &  & & &  &  & & & & 1 &   1\\     
{\bf None}     &  6& &  & & &  & &1& 5&  & & & & 16&  28\\  
{\bf HII/SF}   &   & &  & & &  & & &  &  & &1& &  3&   4\\
{\bf PN }      &   & &  & & &  & & & 1&  & & & &   &   1\\
{\bf post-AGB} &   & &  & & &  & & &  &  &1& & &   &   1\\
\hline
{\bf TOTAL }   & 82&7&39&6&3&15&2&1&20& 1&1&1&1& 28&207\\ 
\enddata
\end{deluxetable}
%\end{document}

\clearpage

%\documentclass[12pt,preprint]{aastex}

%% manuscript produces a one-column, double-spaced document:

%\documentclass[manuscript]{aastex}

%% preprint2 produces a double-column, single-spaced document:

%\documentclass[preprint2]{aastex}

%\begin{document}
 
\begin{deluxetable}{crrrrrrrrr|r}
\tabletypesize{\scriptsize}
\tablecaption{IRAS vs. OPTICAL SPECTRAL TYPES\label{tbl-3}}
\tablewidth{0pt}
\tablehead{
\colhead{} &
\multicolumn{9}{c}{\bf IRAS LRS TYPE}\\
\colhead{\bf OPTICAL}&
\colhead{\bf A} & \colhead{\bf C}   & \colhead{\bf E}   &
\colhead{\bf F} &
\colhead{\bf S} &
\colhead{\bf H} &
\colhead{\bf I} &
\colhead{\bf P} &
\colhead{\bf U} &
\colhead{\bf TOTAL} \\
\colhead{\bf TYPE}&
\\
}
\startdata
{\bf M }      &1& 1& 91&17&19& &1& & &130\\
{\bf S }      & &  &  4&  &  & & & & &  4\\
{\bf C }      & &29&   &  & 1& & & & & 30\\
{\bf K }      & &  &   & 1& 6& & & &1&  8\\
{\bf F }      & &  &   &  & 1& & & & &  1\\
{\bf None}    & &16& 11&  &  & & & &1& 28\\
{\bf H~II/SF} & &  &   &  &  &3& &1& &  4\\
{\bf PN }     & &  &   &  &  &1& & & &  1\\
{\bf post-AGB}& &  &   &  &  & & &1& &  1\\
\hline
{\bf TOTAL }  &1&46&106&18&27&4&1&2&2&207\\

\enddata
\end{deluxetable}
%\end{document}

\clearpage

%\documentclass[12pt,preprint]{aastex}

%% manuscript produces a one-column, double-spaced document:

%\documentclass[manuscript]{aastex}

%% preprint2 produces a double-column, single-spaced document:

%\documentclass[preprint2]{aastex}

%\begin{document}
 
\begin{deluxetable}{crrrrrrrrrrrrrr|r}
\tabletypesize{\scriptsize}
\tablecaption{VARIABILITY TYPES vs. IRAS LRS SPECTRAL TYPES\label{tbl-4}}
\tablewidth{0pt}
\tablehead{
\colhead{} &
\multicolumn{14}{c}{\bf VARIABILITY TYPE}\\
\colhead{\bf LRS}&
\colhead{\bf Mira} & \colhead{\bf SRa}   & \colhead{\bf SRb}   &
\colhead{\bf SRc} &
\colhead{\bf SR} &
\colhead{\bf Lb} &
\colhead{\bf Lc} &
\colhead{\bf L} &
\colhead{\bf variable} &
\colhead{\bf Z+S} &
\colhead{\bf IN} &
\colhead{\bf R} &
\colhead{\bf E} &
\colhead{\bf none} &
\colhead{\bf TOTAL} \\
\colhead{\bf TYPE}&
\\
}
\startdata
{\bf A }       &   & &  & & &  & & & 1&  & & & &   &   1\\      
{\bf C }       & 19&2& 5& &1& 3& & & 3&  & & & & 13&  46\\     
{\bf E }       & 51&5&20&3&2& 5& &1& 8& 1& & & & 10& 106\\     
{\bf F }       & 11& & 5&1& & 1& & &  &  & & & &   &  18\\
{\bf S }       &   & & 8&2& & 6&2& & 6&  & & &1&  2&  27\\     
{\bf H}        &   & &  & & &  & & & 1&  & & & &  3&   4\\  
{\bf I}        &   & & 1& & &  & & &  &  & & & &   &   1\\
{\bf P}        &   & &  & & &  & & &  &  &1&1& &   &   2\\
{\bf U}        &  1& &  & & &  & & & 1&  & & & &   &   2\\
\hline
{\bf TOTAL }   & 82&7&39&6&3&15&2&1&20& 1&1&1&1& 28&207\\ 
\enddata
\end{deluxetable}
%\end{document}

\clearpage

%% manuscript produces a one-column, double-spaced document:

%% preprint2 produces a double-column, single-spaced document:

\begin{deluxetable}{ccccccccccccccccccccc}
\tabletypesize{\scriptsize}
\tablecaption{DIRBE PHOTOMETRY\label{tbl-5}}
\tablewidth{0pt}
\tablehead{
\colhead{IRAS }&
\colhead{Other} & 
\multicolumn{6}{c}{1.25 $\mu$m}&
\multicolumn{6}{c}{2.2 $\mu$m}\\
\colhead{Name}&
\colhead{Name} & 
\colhead{F$_{\nu}$}   & \colhead{$\sigma$}   &
\colhead{$<$err$>$}&
\colhead{$\Delta$mag} &
\colhead{$\sigma$$\Delta$mag} &
\colhead{N}&
\colhead{F$_{\nu}$}   & \colhead{$\sigma$}   &
\colhead{$<$err$>$}&
\colhead{$\Delta$mag} &
\colhead{$\sigma$$\Delta$mag} &
\colhead{N}
\\
\hline
\hline
\multicolumn{14}{c}
{
Available in the Electronic Astronomical Journal at http://www.journals.uchicago.edu/AJ
}
\\
}
\startdata
\enddata
\end{deluxetable}

\clearpage

%\documentclass[12pt,preprint]{aastex}

%% manuscript produces a one-column, double-spaced document:

%\documentclass[manuscript]{aastex}

%% preprint2 produces a double-column, single-spaced document:

%\documentclass[preprint2]{aastex}

%\begin{document}
 
\begin{deluxetable}{cccc|cr|cr|cr|cr|cr|}
%\tabletypesize{\scriptsize}
\tabletypesize{\tiny}
\tablecaption{MEAN DIRBE COLORS\label{tbl-6}}
\tablewidth{0pt}
\tablehead{
\colhead{Variability}&
\colhead{Spectral} & 
\colhead{V$-$$[$12$]$} & 
\colhead{Lum} & 
\multicolumn{2}{c}{2.5 LOG}   &
\multicolumn{2}{c}{2.5 LOG} &
\multicolumn{2}{c}{2.5 LOG} &
\multicolumn{2}{c}{2.5 LOG} &
\multicolumn{2}{c}{2.5 LOG} 
\\
\colhead{Type}&
\colhead{Type} &
\colhead{Limit} &
\colhead{Class} & 
\multicolumn{2}{c}{F$_{1.25}$/F$_{2.2}$}   &
\multicolumn{2}{c}{F$_{2.2}$/F$_{3.5}$} &
\multicolumn{2}{c}{F$_{3.5}$/F$_{4.9}$} &
\multicolumn{2}{c}{F$_{4.9}$/F$_{12}$} &
\multicolumn{2}{c}{F$_{12}$/F$_{25}$}
\\
}
\startdata
      Mira &       OH/IR &       &       & -1.19 $\pm$   0.46 &   9 & -0.38 $\pm$   0.68 &   9 &  0.06 $\pm$   0.31 &  15 & -0.33 $\pm$   0.45 &  16 &  0.04 $\pm$   0.35 &  14\\
      Mira &      oxygen &       &       & -0.59 $\pm$   0.28 &  27 &  0.26 $\pm$   0.29 &  25 &  0.38 $\pm$   0.19 &  33 &  0.15 $\pm$   0.70 &  33 &  0.24 $\pm$   0.31 &  20\\
      Mira &      carbon &       &       & -1.50 $\pm$   0.07 &   2 & -0.87 $\pm$   0.77 &   9 & -0.53 $\pm$   0.56 &  14 & -0.14 $\pm$   0.70 &  12 &  0.46 $\pm$   0.27 &  12\\
      Mira &           S &       &       & -0.87 $\pm$   0.11 &   2 &  0.02 &   1 & -0.08 $\pm$   0.02 &   2 & -0.48        &   1 &  0.20 $\pm$   0.22 &   3\\
     SRa/b &       OH/IR &       &       & -0.90 $\pm$   0.32 &   2 &  0.12 $\pm$   0.24 &   2 &  0.24 $\pm$   0.17 &   2 & -0.06 $\pm$   0.70 &   2 &  0.19 $\pm$   0.27 &   3\\
       SRa &      oxygen &       &       & -0.51 $\pm$   0.14 &   2 &  0.34 &   1 &  0.60 $\pm$   0.20 &   2 &  0.42 $\pm$   0.14 &   2 &            &   0\\
       SRa &      oxygen &       & I/II  &            &   0 &            &   0 &  0.60 &   1 &  0.74        &   1 &  0.52        &   1\\
       SRa &      carbon &       &       & -1.25 &   1 & -0.17        &   1 & -0.31 $\pm$   0.54 &   2 &  0.03 $\pm$   0.20 &   2 &            &   0\\
    SR/SRb &      oxygen &       &       & -0.25 $\pm$   0.14 &  31 &  0.56 $\pm$   0.30 &  29 &  0.71 $\pm$   0.23 &  32 &  0.73 $\pm$   0.43 &  33 &  0.33 $\pm$   0.14 &  14\\
       SRb &      oxygen &       & I/II  & -0.08 &   1 &            &   0 &            &   0 &            &   0 &            &   0\\
       SRb &      carbon &       &       & -0.62 $\pm$   0.20 &   5 &  0.37 $\pm$   0.06 &   4 &  0.89 $\pm$   0.20 &   5 &  0.71 $\pm$   0.15 &   5 &            &   0\\
       SRb &           S &       &       & -0.36 &   1 &  0.48        &   1 &  0.87        &   1 &            &   0 &  0.40        &   1\\
       SRc &      oxygen &       &       & -0.15 $\pm$   0.11 &   6 &  0.64 $\pm$   0.04 &   6 &  0.80 $\pm$   0.10 &   4 &  0.64 $\pm$   0.34 &   4 &  0.28 $\pm$   0.11 &   2\\
      L/Lb &      oxygen &       &       & -0.22 $\pm$   0.31 &   8 &  0.62 $\pm$   0.15 &   9 &  0.76 $\pm$   0.17 &  11 &  0.81 $\pm$   0.70 &   9 &  0.61 $\pm$   0.49 &   3\\
        Lb &      carbon &       &       & -0.58 $\pm$   0.02 &   2 &  0.38 &   1 &  0.61 $\pm$   0.94 &   3 &  0.94 $\pm$   0.35 &   3 &            &   0\\
        Lc &      oxygen &       &       & -0.07 &   1 &  0.72        &   1 &  0.92        &   1 &  1.40        &   1 &  0.89 $\pm$   0.06 &   2\\
  Variable &       OH/IR &       &       &            &   0 &            &   0 & -0.75 $\pm$   0.30 &   2 & -0.78 $\pm$   0.49 &   3 & -0.74 $\pm$   0.28 &   2\\
  Variable &       OH/IR &       & I/II  &            &   0 &            &   0 &            &   0 &            &   0 &            &   0\\
  Variable &      oxygen &  $<$5 &       &  0.14 $\pm$   0.11 &   5 &  0.76 $\pm$   0.02 &   5 &  0.87 $\pm$   0.03 &   5 &  1.51 $\pm$   0.04 &   5 &  1.02 &   1\\
  Variable &      oxygen & $>$12 &       & -0.56 $\pm$   0.52 &   2 &  0.36 $\pm$   0.25 &   3 &  0.56 $\pm$   0.27 &   4 & -0.09 $\pm$   0.34 &   5 &  0.33 &   1\\
  Variable &      oxygen &       & I/II  &            &   0 &            &   0 &            &   0 &            &   0 &  0.40 &   1\\
  Variable &      carbon &       &       &            &   0 &            &   0 & -0.63 &   1 & -0.26 $\pm$   0.86 &   3 &  0.37 $\pm$   0.45 &   2\\
      None &      oxygen &  $<$5 &       &  0.14 $\pm$   0.01 &   2 &  0.72 $\pm$   0.04 &   2 &  0.95 $\pm$   0.07 &   2 &  1.44 $\pm$   0.09 &   2 &            &   0\\
      None &      oxygen & $>$12 &       &            &   0 &            &   0 & -0.14 $\pm$   0.12 &   2 & -0.61 $\pm$   0.30 &  10 & -0.07 $\pm$   0.14 &   7\\
      None &      oxygen &       & I/II  &  0.83 &   1 &  0.84        &   1 &  0.65        &   1 &  1.75        &   1 &            &   0\\
      None &      carbon &       &       &            &   0 &            &   0 & -0.89 $\pm$   0.26 &   3 & -0.27 $\pm$   0.55 &  10 &  0.33 $\pm$   0.16 &   7\\
\enddata
\end{deluxetable}

%\end{document}

\clearpage

%\documentclass[12pt,preprint]{aastex}

%% manuscript produces a one-column, double-spaced document:

%\documentclass[manuscript]{aastex}

%% preprint2 produces a double-column, single-spaced document:

%\documentclass[preprint2]{aastex}

%\begin{document}
 
\begin{deluxetable}{cccc|ccc|ccc|ccc|ccc|ccc|ccc|ccc|ccc}
%\tabletypesize{\scriptsize}
\tabletypesize{\tiny}
\tablecaption{MEAN DIRBE AMPLITUDES OF VARIATION\label{tbl-7}}
\tablewidth{0pt}
\tablehead{
\colhead{Var}&
\colhead{Spectral} & 
\colhead{V-$[$12$]$} & 
\colhead{Lum} & 
\multicolumn{3}{c}{1.25 {$\mu$}m }&
\multicolumn{3}{c}{2.2 {$\mu$}m }&
\multicolumn{3}{c}{3.5 {$\mu$}m }&
\multicolumn{3}{c}{4.9 {$\mu$}m }&
\multicolumn{3}{c}{12 {$\mu$}m }&
\multicolumn{3}{c}{25 {$\mu$}m }
\\
\colhead{Type}&
\colhead{Type} & 
\colhead{Limit}&
\colhead{Class} & 
\colhead{$<$${\Delta}$mag$>$}   &
\colhead{$<$${\sigma}$$>$}   &
\colhead{N} &
\colhead{$<$${\Delta}$mag$>$}   &
\colhead{$<$${\sigma}$$>$}   &
\colhead{N} &
\colhead{$<$${\Delta}$mag$>$}   &
\colhead{$<$${\sigma}$$>$}   &
\colhead{N} &
\colhead{$<$${\Delta}$mag$>$}   &
\colhead{$<$${\sigma}$$>$}   &
\colhead{N} &
\colhead{$<$${\Delta}$mag$>$}   &
\colhead{$<$${\sigma}$$>$}   &
\colhead{N} &
\colhead{$<$${\Delta}$mag$>$}   &
\colhead{$<$${\sigma}$$>$}   &
\colhead{N} 
\\
}
\startdata

      Mira &       OH/IR &      &     &  1.46$\pm$  0.65&  0.10&         4 &  0.60$\pm$  0.38&  0.10&        12 &  0.68$\pm$  0.31&  0.07&        15 &  0.62$\pm$  0.30&  0.06&        16 &  0.71$\pm$  0.25&  0.09&        15 &  0.58$\pm$  0.22&  0.11&        10\\
      Mira &      oxygen &      &     &  0.90$\pm$  0.44&  0.07&        26 &  0.53$\pm$  0.30&  0.03&        27 &  0.53$\pm$  0.23&  0.05&        33 &  0.47$\pm$  0.27&  0.04&        34 &  0.58$\pm$  0.26&  0.12&        35 &  0.62$\pm$  0.21&  0.14&        15\\
      Mira &      carbon &      &     &  0.92      &  0.11&         1 &  0.67$\pm$  0.44&  0.07&        10 &  0.81$\pm$  0.37&  0.07&        12 &  0.74$\pm$  0.36&  0.05&        13 &  0.54$\pm$  0.21&  0.10&        18 &  0.55$\pm$  0.18&  0.15&         8\\
      Mira &           S &      &     &  1.19$\pm$  0.27&  0.15&         2 &  0.21      &  0.07&         1 &  0.99$\pm$  0.20&  0.05&         2 &  1.00      &  0.04&         1 &  0.91$\pm$  0.08&  0.08&         3 &  0.90$\pm$  0.06&  0.19&         3\\
     SRa/b &       OH/IR &      &     &  0.80$\pm$  0.72&  0.08&         3 &  0.68$\pm$  0.39&  0.03&         2 &  0.43$\pm$  0.34&  0.03&         3 &  0.59$\pm$  0.12&  0.03&         2 &  0.39$\pm$  0.30&  0.07&         3 &  0.42$\pm$  0.29&  0.15&         3\\
       SRa &      oxygen &      &     &  0.36$\pm$  0.45&  0.04&         2 &  0.50      &  0.01&         1 &  0.21$\pm$  0.19&  0.04&         2 &  0.11$\pm$  0.10&  0.05&         2 &  0.21$\pm$  0.06&  0.10&         2 &        &  0.15&         0\\
       SRa &      oxygen &      &I/II &        &  0.04&         0 &        &  0.01&         0 &  0.30      &  0.15&         1 &  0.09      &  0.01&         1 &  0.08      &  0.06&         1 &  0.22      &  0.17&         1\\
       SRa &      carbon &      &     &  0.46      &  0.05&         1 &  0.22      &  0.01&         1 &  0.33$\pm$  0.22&  0.03&         2 &  0.29$\pm$  0.11&  0.01&         2 &  0.20$\pm$  0.06&  0.03&         2 &        &  0.17&         0\\
       SRb &      oxygen &      &     &  0.15$\pm$  0.09&  0.03&        29 &  0.12$\pm$  0.21&  0.02&        31 &  0.14$\pm$  0.18&  0.03&        33 &  0.10$\pm$  0.16&  0.02&        33 &  0.16$\pm$  0.11&  0.10&        34 &  0.21$\pm$  0.07&  0.17&        13\\
       SRb &     oxygen  &      &I/II &  0.03      &  0.02&         1 &  0.01      &  0.01&         1 &  0.05      &  0.02&         1 &        &  0.02&         0 &  0.14      &  0.11&         1 &        &  0.17&         0\\
       SRb &      carbon &      &     &  0.28$\pm$  0.24&  0.05&         5 &  0.11$\pm$  0.13&  0.02&         5 &  0.13$\pm$  0.09&  0.03&         5 &  0.12$\pm$  0.10&  0.03&         5 &  0.19$\pm$  0.04&  0.16&         5 &        &  0.17&         0\\
       SRb &           S &      &     &  0.10      &  0.03&         1 &  0.01      &  0.01&         1 &  0.04      &  0.02&         1 &        &  0.03&         0 &  0.04      &  0.04&         1 &  0.11      &  0.09&         1\\
       SRc &      oxygen &      &     &  0.09$\pm$  0.04&  0.03&         6 &  0.06$\pm$  0.02&  0.01&         6 &  0.07$\pm$  0.03&  0.03&         6 &  0.06$\pm$  0.02&  0.02&         4 &  0.12$\pm$  0.05&  0.09&         6 &  0.18$\pm$  0.04&  0.18&         2\\
        Lb &      oxygen &      &     &  0.08$\pm$  0.06&  0.03&         8 &  0.09$\pm$  0.13&  0.02&         9 &  0.10$\pm$  0.09&  0.04&        11 &  0.07$\pm$  0.08&  0.03&        11 &  0.19$\pm$  0.14&  0.14&        10 &  0.19$\pm$  0.07&  0.20&         2\\
        Lb &      carbon &      &     &  0.15$\pm$  0.05&  0.09&         2 &  0.10      &  0.02&         1 &  0.09$\pm$  0.02&  0.06&         2 &  0.12$\pm$  0.13&  0.04&         4 &  0.24$\pm$  0.09&  0.20&         3 &        &  0.20&         0\\
        Lc &      oxygen &      &     &  0.05$\pm$  0.01&  0.03&         2 &  0.02      &  0.01&         1 &  0.04$\pm$  0.00&  0.02&         2 &  0.02      &  0.01&         1 &  0.11$\pm$  0.07&  0.04&         2 &  0.25$\pm$  0.14&  0.14&         2\\
  variable &       OH/IR &      &     &        &  0.03&         0 &        &  0.01&         0 &  0.45      &  0.18&         1 &  0.88$\pm$  0.52&  0.10&         3 &  0.76$\pm$  0.48&  0.11&         3 &  0.83$\pm$  0.23&  0.12&         2\\
  variable &       OH/IR &      &I/II &        &  0.03&         0 &        &  0.01&         0 &  0.16      &  0.02&         1 &        &  0.10&         0 &  0.11      &  0.01&         1 &        &  0.12&         0\\
  variable &      oxygen & $<$5 &     &  0.04$\pm$  0.01&  0.03&         6 &  0.01$\pm$  0.00&  0.01&         5 &  0.03$\pm$  0.01&  0.02&         6 &  0.01$\pm$  0.00&  0.02&         5 &  0.21$\pm$  0.16&  0.15&         5 &  0.27$\pm$  0.09&  0.17&         2\\
  variable &      oxygen &$>$12 &     &  0.12$\pm$  0.04&  0.04&         2 &  0.51$\pm$  0.35&  0.06&         4 &  0.43$\pm$  0.37&  0.09&         4 &  0.39$\pm$  0.30&  0.06&         5 &  0.55$\pm$  0.36&  0.19&         4 &        &  0.17&         0\\
  variable &      oxygen &      &I/II &        &  0.04&         0 &        &  0.06&         0 &        &  0.09&         0 &        &  0.06&         0 &  0.07      &  0.05&         1 &  0.19      &  0.12&         1\\
  variable &      carbon &      &     &        &  0.04&         0 &        &  0.06&         0 &        &  0.09&         0 &  1.08$\pm$  0.29&  0.08&         3 &  0.76$\pm$  0.16&  0.11&         3 &  0.77      &  0.18&         1\\
      none &      oxygen & $<$5 &     &  0.04$\pm$  0.01&  0.03&         2 &  0.01$\pm$  0.01&  0.01&         2 &  0.02$\pm$  0.00&  0.03&         2 &  0.03$\pm$  0.01&  0.02&         2 &  0.17$\pm$  0.01&  0.14&         2 &        &  0.18&         0\\
      none &      oxygen &$>$12 &     &        &  0.03&         0 &        &  0.01&         0 &  0.63      &  0.05&         1 &  0.72$\pm$  0.28&  0.11&         9 &  0.65$\pm$  0.29&  0.13&         7 &  0.41$\pm$  0.13&  0.15&         4\\
      none &      oxygen &      &I/II &  0.03      &  0.03&         1 &  0.01      &  0.01&         1 &  0.03      &  0.03&         1 &  0.02      &  0.02&         1 &  0.20      &  0.17&         1 &        &  0.15&         0\\
      none &      carbon &      &     &        &  0.03&         0 &        &  0.01&         0 &  1.16$\pm$  0.05&  0.12&         2 &  0.87$\pm$  0.27&  0.07&        13 &  0.73$\pm$  0.13&  0.12&        11 &  0.67$\pm$  0.11&  0.15&         6\\
\enddata
\end{deluxetable}

%\end{document}

\clearpage
  
\begin{deluxetable}{cccc|cc|cc|cc|cc|cc|cc|cc|cc}
%\tabletypesize{\scriptsize}
\tabletypesize{\tiny}
\tablecaption{MEAN AMPLITUDE RATIOS\label{tbl-8}}
\tablewidth{0pt}
\tablehead{
\colhead{Variability}&
\colhead{Spectral} & 
\colhead{V-$[$12$]$} & 
\colhead{Lum} & 
\multicolumn{2}{c}{1.25/2.2}&
\multicolumn{2}{c}{2.2/3.5 }&
\multicolumn{2}{c}{3.5/4.9 }&
\multicolumn{2}{c}{4.9/12}&
\multicolumn{2}{c}{12/25}&
\multicolumn{2}{c}{25/60 }
\\
\colhead{Type}&
\colhead{Type} & 
\colhead{Limit}&
\colhead{Class} & 
\colhead{${\Delta}$mag/${\Delta}$mag}   &
\colhead{N} &
\colhead{${\Delta}$mag/${\Delta}$mag}   &
\colhead{N} &
\colhead{${\Delta}$mag/${\Delta}$mag}   &
\colhead{N} &
\colhead{${\Delta}$mag/${\Delta}$mag}   &
\colhead{N} &
\colhead{${\Delta}$mag/${\Delta}$mag}   &
\colhead{N} &
\colhead{${\Delta}$mag/${\Delta}$mag}   &
\colhead{N} 
\\
}
\startdata

      Mira &       OH/IR &      &     &  1.88$\pm$  0.27&         3 &  1.19$\pm$  0.07&         6 &  1.08$\pm$  0.20&        12 &  0.99$\pm$  0.11&         8 &  1.22$\pm$  0.06&         6 &           &         0\\
      Mira &      oxygen &      &     &  1.99$\pm$  0.69&        21 &  1.09$\pm$  0.32&        24 &  1.20$\pm$  0.26&        28 &  0.98$\pm$  0.25&        15 &  1.02$\pm$  0.14&         6 &           &         0\\
      Mira &      carbon &      &     &  2.79      &         1 &  0.92$\pm$  0.24&         7 &  1.22$\pm$  0.24&         8 &  1.28$\pm$  0.18&         8 &  1.17$\pm$  0.07&         3 &  1.37      &         1\\
      Mira &           S &      &     &           &         0 &           &         0 &  0.85      &         1 &  1.01      &         1 &  0.98      &         1 &           &         0\\
     SRa/b &       OH/IR &      &     &  1.73$\pm$  0.14&         2 &  1.08$\pm$  0.24&         2 &  0.99$\pm$  0.18&         2 &  1.06$\pm$  0.08&         2 &           &         0 &           &         0\\
       SRa &      oxygen &      &     &  1.36      &         1 &  1.47      &         1 &  1.89      &         1 &  1.12      &         1 &           &         0 &           &         0\\
       SRa &      oxygen &      &I/II &           &         0 &           &         0 &           &         0 &           &         0 &           &         0 &           &         0\\
       SRa &      carbon &      &     &  2.09      &         1 &  1.29      &         1 &  1.05$\pm$  0.34&         2 &  1.43$\pm$  0.16&         2 &           &         0 &           &         0\\
       SRb &      oxygen &      &     &  1.89$\pm$  0.73&        11 &  0.84$\pm$  0.21&        10 &  1.45$\pm$  0.34&         9 &  1.39$\pm$  0.26&         2 &           &         0 &           &         0\\
       SRb &     oxygen  &      &I/II &           &         0 &           &         0 &           &         0 &           &         0 &           &         0 &           &         0\\
       SRb &      carbon &      &     &  2.71$\pm$  0.97&         2 &  1.17      &         1 &  1.00      &         1 &           &         0 &           &         0 &           &         0\\
       SRb &           S &      &     &           &         0 &           &         0 &           &         0 &           &         0 &           &         0 &           &         0\\
       SRc &      oxygen &      &     &           &         0 &           &         0 &           &         0 &  0.90      &         1 &           &         0 &           &         0\\
        Lb &      oxygen &      &     &  1.65$\pm$  0.33&         2 &           &         0 &           &         0 &           &         0 &           &         0 &           &         0\\
        Lb &      carbon &      &     &           &         0 &           &         0 &           &         0 &           &         0 &           &         0 &           &         0\\
        Lc &      oxygen &      &     &           &         0 &           &         0 &           &         0 &           &         0 &           &         0 &           &         0\\
       var &       OH/IR &      &     &           &         0 &           &         0 &           &         0 &  1.17$\pm$  0.16&         2 &  1.23$\pm$  0.03&         2 &           &         0\\
       var &       OH/IR &      &I/II &           &         0 &           &         0 &           &         0 &           &         0 &           &         0 &           &         0\\
       var &      oxygen & $<$5 &     &           &         0 &           &         0 &           &         0 &           &         0 &           &         0 &           &         0\\
       var &      oxygen &$>$12 &     &           &         0 &  1.09      &         1 &  1.07      &         1 &  0.86      &         1 &           &         0 &           &         0\\
       var &      oxygen &      &I/II &           &         0 &           &         0 &           &         0 &           &         0 &           &         0 &           &         0\\
       var &      carbon &      &     &           &         0 &           &         0 &           &         0 &  1.41$\pm$  0.13&         2 &           &         0 &           &         0\\
      none &      oxygen & $<$5 &     &           &         0 &           &         0 &           &         0 &           &         0 &           &         0 &           &         0\\
      none &      oxygen &$>$12 &     &           &         0 &           &         0 &  1.15      &         1 &  1.20$\pm$  0.24&         3 &  1.20$\pm$  0.09&         2 &           &         0\\
      none &      oxygen &      &I/II &           &         0 &           &         0 &           &         0 &           &         0 &           &         0 &           &         0\\
      none &      carbon &      &     &           &         0 &           &         0 &  1.19$\pm$  0.05&         2 &  1.44$\pm$  0.16&         6 &  1.11$\pm$  0.08&         3 &           &         0\\
\enddata
\end{deluxetable}

\end{document}